\documentclass[12pt,preprint]{aastex}

\usepackage{graphicx, color}
\usepackage{dcolumn}
\usepackage{bm}
\usepackage{booktabs}
\usepackage{natbib}
\usepackage{lscape}

\newcommand {\vect}[1]{\mbox{\boldmath $#1$}}

\newcommand {\dif}[3][]{\frac{d^{#1}#2}{d#3^{#1}}}
\newcommand {\pdif}[3][]{\frac{\partial^{#1}#2}{\partial#3^{#1}}}

\newcommand {\lsim}{\hspace{0.3em}\raisebox{0.4ex}{$<$}\hspace{-0.75em}\raisebox{-.7ex}{$\sim$}\hspace{0.3em}}
\newcommand {\gsim}{\hspace{0.3em}\raisebox{0.4ex}{$>$}\hspace{-0.75em}\raisebox{-.7ex}{$\sim$}\hspace{0.3em}}

\makeatletter
\def\mart{\@ifnextchar[{\mart@@}{\mart@}}
\def\mart@@[#1]#2{\sqrt[#1]{\mathstrut{#2}}}
\def\mart@#1{\sqrt{\mathstrut{#1}}}
\makeatother
\newcommand {\Alfven}{Alfv\'{e}n}
\newcommand {\Elsasser}{Els$\ddot{\rm a}$sser }
\newcommand{\myemail}{minoshim@jamstec.go.jp}

\long\def\symbolfootnote[#1]#2{\begingroup%
\def\thefootnote{\fnsymbol{footnote}}\footnote[#1]{#2}\endgroup}

\begin{document}

\title{Dependence of the saturation level of magnetorotational instability on gas pressure and magnetic Prandtl number}
\shorttitle{Pressure and magnetic Prandtl number dependence of MRI}
\author{Takashi Minoshima\altaffilmark{1}, Shigenobu Hirose\altaffilmark{1}, and Takayoshi Sano\altaffilmark{2}}
\altaffiltext{1}{Department of Mathematical Science and Advanced Technology, Japan Agency for Marine-Earth Science and Technology, 3173-25, Syowa-machi, Kanazawaku, Yokohama 236-0001, Japan}
\altaffiltext{2}{Institute of Laser Engineering, Osaka University, Suita, Osaka 565-0871, Japan}
\shortauthors{Minoshima et al.}
\email{\myemail}

\begin{abstract}
A large set of numerical simulations of magnetohydrodynamic (MHD) turbulence induced by the magnetorotational instability (MRI) is presented.
{ Revisiting the previous survey conducted by \cite{2004ApJ...605..321S}, we investigate the gas pressure dependence of the saturation level.
In ideal MHD simulations, the gas pressure dependence is found to be very sensitive to the choice of a numerical scheme.
This is because the numerical magnetic Prandtl number varies according to the scheme as well as the pressure, which considerably affects the results.}
The saturation level is more sensitive to the numerical magnetic Prandtl number than the pressure. 
{ In MHD simulations with explicit viscosity and resistivity, the saturation level increases with the physical magnetic Prandtl number, and it is almost independent of the gas pressure when the magnetic Prandtl number is constant.}
This is indicative of the incompressible turbulence saturated by the secondary tearing instability.

\end{abstract}


\section{INTRODUCTION}\label{sec:introduction}
Space is filled with dilute, magnetized plasmas.
Plasmas that exhibit sufficiently large Reynolds and magnetic Reynolds numbers are expected to be in a turbulent state.
For example, weakly magnetized accretion disks are subject to magnetic turbulence induced by the magnetorotational instability \cite[MRI;][]{1991ApJ...376..214B,1998RvMP...70....1B}.
Since numerical simulations have revealed that the Maxwell stress in the MRI-induced turbulence is orders of magnitude larger than the viscous stress estimated from ordinary molecular viscosity, the MRI-induced turbulence is believed to play a critical role in the outward angular momentum transport and subsequent mass accretion toward the central star \citep{1995ApJ...440..742H,1996ApJ...464..690H}.
The nonlinear stage of turbulence in magnetized plasmas has been extensively studied via magnetohydrodynamic (MHD) simulations.


Ideal MHD equations without physical viscosity and resistivity are often utilized for the numerical study of astrophysical plasma dynamics.
Artificial viscosity and resistivity should be explicitly added to central (e.g., Lax-Wendroff) or operator-splitting (e.g., ZEUS by \cite{1992ApJS...80..753S,1992ApJS...80..791S}) schemes to suppress spurious oscillations and capture shocks.
On the other hand, Godunov-type schemes {automatically embed} sufficient numerical diffusion because of their upwind property by virtue of a Riemann solver.
Recent Godunov-type schemes such as NIRVANA \citep{2004JCoPh.196..393Z,2008CoPhC.179..227Z}, RAMSES \citep{2006A&A...457..371F}, PLUTO \citep{2007ApJS..170..228M}, and ATHENA \citep{2008ApJS..178..137S} have been designed to capture shocks and discontinuities via exact or approximate Riemann solvers, and to resolve small-scale turbulence via high-order reconstruction techniques.
The spatial {accuracies} of the fluid and magnetic field in Godunov-type schemes are not necessarily identical because of the treatment of the induction equation (divergence or curl form) and special consideration given to the divergence-free condition for the magnetic field.
\cite{2011ApJ...737...13K} have compared the {accuracy} of nine MHD simulation codes from a turbulence decay simulation, and they have shown that a code with high {accuracy} for velocity does not necessarily possess high {accuracy} for magnetic field, and vice versa.
The numerical magnetic Prandtl number that is defined as the ratio of the numerical viscosity to resistivity varies according to the applied schemes, and it can be smaller or larger than unity.

The magnetic Prandtl number $Pr_{m}$ determines and controls essential properties of MHD turbulence.
The number relates the viscous dissipation scale $l_{\nu}$ to the resistive dissipation scale $l_{\eta}$.
For example, \cite{2004ApJ...612..276S} have argued that the kinetic energy input at the system scale cascades down to the Kolmogorov dissipation scale $l_{\nu}$, and subsequently, viscous eddies amplify small-scale magnetic fields down to $l_{\eta} \sim Pr_{m}^{-1/2} l_{\nu}$ for $Pr_{m} > 1$.
When Spitzer values are used for the viscosity and resistivity of fully ionized, weakly magnetized plasmas, the magnetic Prandtl number is estimated to be $Pr_{m} \sim 10^{-5} T^4/n$, where $T$ and $n$ denote the temperature and number density in cgs units, respectively \citep{1962pfig.book.....S}.
Astrophysical objects can have magnetic Prandtl numbers that lie far from $Pr_{m} = 1$.
There are extreme environments that have $Pr_{m} \gg 1$ (e.g., stellar corona, active galactic nuclei disks) or $Pr_{m} \ll 1$ (e.g., stellar interior, young stellar object disks).
In such cases, there is a very wide sub-viscous or sub-resistive range over which the fluid and magnetic field are decoupled with each other.
It can considerably alter the generation and dissipation processes of MHD turbulence with respect to those observed for $Pr_{m} = 1$.
\cite{2004ApJ...612..276S} have performed numerical simulations of forced MHD turbulence, and they have shown that a small-scale dynamo with $Pr_{m} \gg 1$ generates magnetic fields in the sub-viscous range.
The magnetic field energy positively correlates with $Pr_{m}$ and it does not satisfy scale-by-scale equipartition with the kinetic energy in the saturated state. 
Such physics is not included in ideal MHD simulations, whose numerical magnetic Prandtl number is of the order of unity.

When ideal MHD equations are solved with a Godunov-type scheme to study MHD turbulence, the viscous and resistive dissipation scales and hence the magnetic Prandtl number are implicitly determined depending on the choice of the reconstruction technique and the Riemann solver.
It is of critical importance to clarify effects of numerical dissipation on practical MHD turbulence problems.
To this end, we conduct a large set of ideal MHD simulations of the MRI.
We revisit an extensive survey of the gas pressure dependence of the saturation level of the MRI-induced turbulence as conducted by \cite{2004ApJ...605..321S}.
Our simulation reveals that the saturation level and its pressure dependence are very sensitive to the numerical magnetic Prandtl number, thereby indicating the necessity of explicit viscosity and resistivity to control the magnetic Prandtl number.
Consequently, we conduct visco-resistive MHD simulations in order to investigate the parameter dependence and discuss the saturation mechanism of the MRI-induced turbulence.

This paper is organized as follows.
In Section \ref{sec:numerical-method}, we describe our numerical method based on a Godunov-type scheme.
In Section \ref{sec:results}, we present simulation results.
First, we assess the effect of numerical dissipation on the saturation level of the MRI-induced turbulence with the ideal MHD simulation in Section \ref{sec:norm-resol-case}.
Subsequently, we investigate the parameter dependence of the saturation level with the visco-resistive MHD simulation in Section \ref{sec:high-resolution-with}.
Section \ref{sec:discussion} discusses the saturation mechanism of the MRI-induced turbulence based on the simulation results.
Finally, we summarize the paper in Section \ref{sec:conclusion}.

\section{NUMERICAL METHOD}\label{sec:numerical-method}
We perform three-dimensional compressible MHD simulations under a local shearing box approximation \citep{1995ApJ...440..742H,2010ApJS..189..142S}. Without considering the viscosity, resistivity, and gravity stratification, the relevant equations are as follows,
\begin{eqnarray}
&&\pdif{\rho}{t} + \nabla \cdot \left(\rho \vect{u}\right) = 0,\label{eq:1}\\
&&\pdif{\rho \vect{u'}}{t} + \nabla \cdot \left[ \rho \vect{u} \vect{u'} + \left(P+\frac{B^2}{8 \pi}\right){\rm I}-\frac{\vect{B}\vect{B}}{4 \pi} \right] = - 2 \vect{\Omega} \times \rho \vect{u'} + q \Omega \rho u_{x}' \vect{e}_y,\label{eq:2}\\
&&\pdif{\vect{B}}{t}+\nabla \times \vect{E} = 0,\label{eq:3}\\
&&\vect{E}=-\vect{u}\times{\vect{B}},\label{eq:5}
\end{eqnarray} 
where $\rho$ denotes the density, $P$ the gas pressure, $\vect{B}$ and $\vect{E}$ the magnetic and electric fields, $\vect{u} = \vect{u_0} + \vect{u'}$ the total velocity, $\vect{u_0} = -q \Omega x \vect{e}_y$ the background flow in the shearing box (local linear expansion of the rotating motion), $\vect{u'}$ the deviation from the background shear flow, $\vect{\Omega}=\Omega \vect{e}_z$ the rotation axis, $q$ the shear parameter ($q=3/2$ for Keplerian disks is used). 
{ The entropy equation is omitted because we adopt an isothermal equation of state, $P/\rho = c_{s}^2 = {\rm constant}$} where $c_{s}$ is the sound speed.

\subsection{Numerical Algorithm}\label{sec:numerical-algorithm}
The equations are advanced based on a technique proposed by \cite{2010ApJS..189..142S}.
Their technique decomposes the equations into the MHD part including the shearing source terms (right-hand-side of equation (\ref{eq:2})) and the orbital advection part with the background shear flow $\vect{u_0}$.
The latter part is advanced as simple one-dimensional linear advection equations because $\vect{u}_0$ has only a $y$-component and is independent of $y$ and time.
The shearing source terms are integrated by means of a second-order Crank$-$Nicholson method so as to guarantee the conservation of epicyclic motion in a discrete sense.

For the MHD part, we adopt a Godunov-type scheme (Minoshima, T., Matsumoto, Y., and Miyoshi, T., in preparation).
The scheme involves (i) nonlinear reconstruction of characteristic variables from cell centers to cell faces, (ii) evaluation of numerical fluxes dimension-by-dimension with a Riemann solver, and (iii) update of the induction equation { with a method similar to} the upwind constrained transport method \citep{2000ApJ...530..508L,2004JCoPh.195...17L,2007A&A...473...11D}.
The induction equation is advanced in curl form to preserve the divergence-free condition for the magnetic field in a discrete form to machine accuracy.
{ Details of this method are described in Appendix \ref{sec:appendix1}.}
For nonlinear reconstruction, we adopt a fifth-order Weighted-ENO scheme \citep[WENO;][]{1996JCoPh.126..202J} or a combination of a fifth-order WENO-Z scheme \citep{2008JCoPh.227.3191B} and a monotonicity-preserving scheme \citep{1997JCoPh.136...83S,2000JCoPh.160..405B}. 
Hereafter, the latter is termed as the WZMP scheme, whose accuracy is slightly better than that of the WENO scheme. 
For Riemann solvers, we adopt the single-state Harten$-$Lax$-$van Leer (HLL) approximate Riemann solver \citep{1983SIAMrev.25..35M} or the multi-state HLLR approximate Riemann solver \citep{2008ASPC..385..279M}. 
The former solver resolves the fast mode, and the latter resolves the {\Alfven} mode as well. 

{ As is described in Appendix \ref{sec:appendix1}}, our simulation code can use different reconstruction functions and Riemann solvers for the fluid and magnetic field.
Consequently, we refer to the applied scheme as F1-R1-F2-R2, where ``F1'' and ``R1'' denote the reconstruction function and Riemann solver for the fluid, and ``F2'' and ``R2'' denote those for the magnetic field, respectively.
For example, the WENO-HLLR-WZMP-HLL scheme uses the WENO reconstruction and the HLLR Riemann solver for the fluid update, and the WZMP reconstruction and the HLL Riemann solver for the magnetic field update.
By varying the combination of the reconstruction function and the Riemann solver for the fluid and magnetic field, we assess { the effect of numerical dissipation on the MRI-induced turbulence.}

\begin{figure}[p]
\centering
\epsscale{1.0}
\plotone{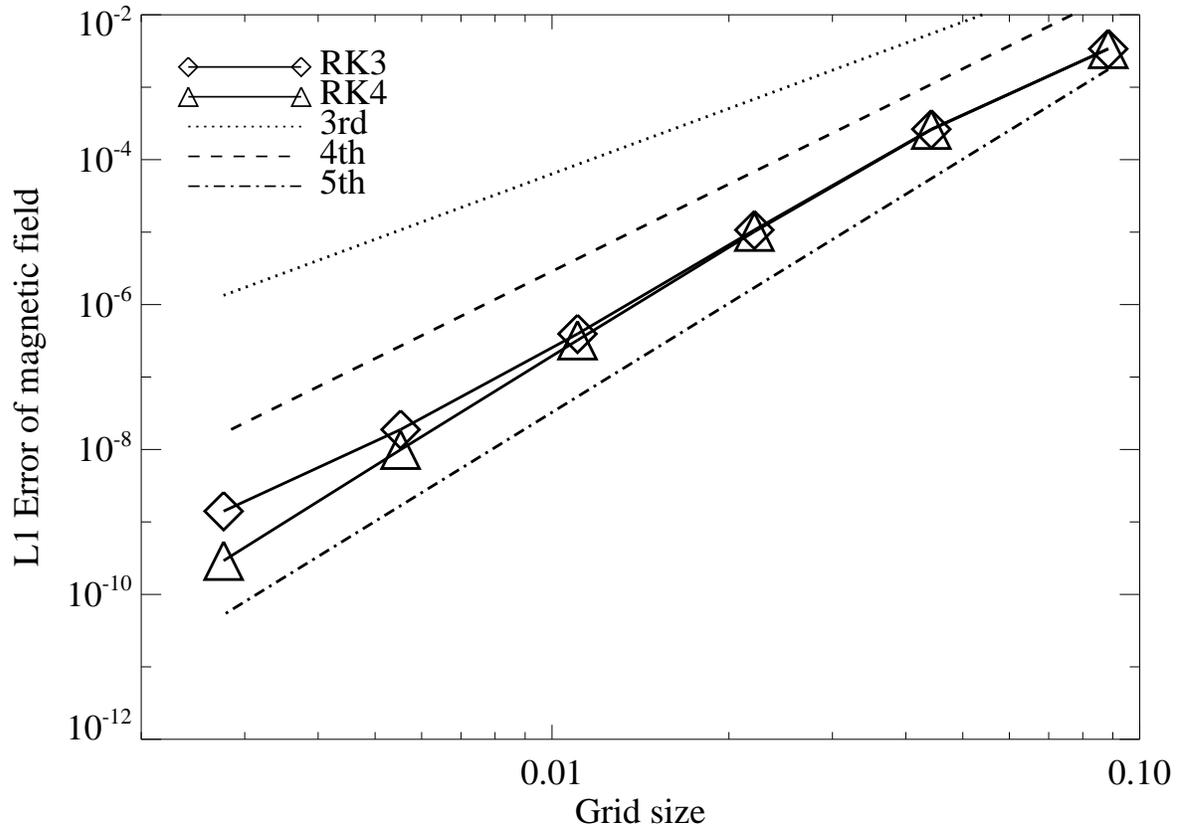}
\caption{$L_1$ error norm of the two-dimensional {\Alfven} wave propagation problem as a function of the grid size. The diamonds and triangles are obtained with the third and fourth order Runge-Kutta time integration, respectively. The dotted, dashed, and dash-dotted lines represent the third, fourth, and fifth order of accuracy, respectively.}
\label{fig:awave}
\end{figure}

{ We adopt a finite volume approach.
Finite volume schemes require reconstruction of the numerical flux over cell faces to achieve a desired order of accuracy in multidimension \citep[e.g.,][]{2004JCoPh.201..238T}. 
We ignore this reconstruction for simplicity, that is, we approximate $\overline{\vect{F}} = \vect{F}(\overline{\vect{U}})$ where $\vect{U},\vect{F}$ are the physical variable and the corresponding flux, and the overline represents the average over the cell face.
This approximation is inaccurate when $\vect{F}$ is a nonlinear function of $\vect{U}$.
It degrades the order of accuracy in multidimension even using a high order reconstruction dimension-by-dimension.

We examine the order of accuracy of our scheme from the propagation of circular-polarized {\Alfven} waves in two-dimensional homogeneous medium.
The propagation angle is $45^\circ$ with respect to the $x$ axis and the plasma beta value is 0.1.
We use the WENO-HLLR-WENO-HLL scheme and the third-order TVD Runge-Kutta time integration \citep{1988JCoPh..77..439S}.
The order of accuracy of the scheme is presented in Figure \ref{fig:awave}.
The scheme preserves fifth order of accuracy at a coarse grid size, but the accuracy becomes worse than the fourth order at a fine grid size.
In this problem, the order of accuracy can be retrieved by increasing the order of the Runge-Kutta time integration, implying that multi-dimensional properties are somewhat incorporated by using the multi-step time integration.
The similar result is obtained with the WZMP-HLLR-WZMP-HLL scheme.
However, this does not necessarily hold for practical problems in highly inhomogeneous medium. 
The order of accuracy may be degraded as much as the second order.
In following simulations, we use the third-order TVD Runge-Kutta time integration.
The fourth-order time integration does not make a significant difference.
}

\cite{2010ApJS..189..142S} have pointed out that the same high-order reconstruction function used in the computational domain is required for the remapping of variables to the shearing boundary to maintain the accuracy of the calculation.
We confirm that a simple arithmetic average 
 at the shearing boundary severely degrades the accuracy of the simulation.
Consequently, we remap the variables by solving the linear advection as follows.
The shearing distance at the left boundary relative to the right, $L_s = q \Omega L t$, is divided into integer and non-integer numbers of grids, $N={\rm int}(L_s/\Delta)$ and $D=L_s/\Delta-N$. 
Subsequently, the variables at the right boundary $U^{R}$ are remapped to the left boundary such that $U^{L}(y_j) = U^{R}(y_{j-N}-D\Delta)$.
We use the same reconstruction function to shift $U^{R}$ with a non-integer distance of $D\Delta$.

The growth of the MRI is very sensitive to the amplitude of the net magnetic flux.
The use of the nonlinear reconstruction and the shearing periodic boundary violates the conservation of the variables \citep{2007CoPhC.176..652G}. 
To avoid the problem, we correct the numerical flux of $B_{z}$ (that is, $E_{y}$) at the shearing boundary $(x=\pm 0.5)$ in a manner similar to that adopted by \cite{2010ApJS..189..142S}.

\subsection{Initial Setup}\label{sec:initial-setup}
The space, time, and velocity are normalized by the system size $L$, the rotation period $\Omega^{-1}$, and the rotation velocity $L\Omega$, respectively.
The simulation domain of $[-0.5,0.5]\times[-2,2]\times[-0.5,0.5]$ is resolved with grid points of  $32\times128\times32$ (grid width $\Delta = 1/32$) in Section \ref{sec:norm-resol-case}.
High-resolution simulations with grid points of $64\times256\times64$ ($\Delta = 1/64$) are also conducted to { check numerical convergence in Section \ref{sec:convergence} and} include explicit viscosity and resistivity in Section \ref{sec:high-resolution-with}.
The boundary condition is periodic along the $y$ (azimuthal) and $z$ (vertical) directions, and shearing periodic along the $x$ (radial) direction, respectively.
The initial condition is uniform, i.e., $(\rho,u_x',u_y',u_z',P,B_x,B_y,B_z)=(1,0,0,0,P_0,0,0,B_0)$, and we impose an incompressible small random noise to $u_x'$ uniformly in the $y-z$ plane and constantly in the $x$ direction to initiate the instability.
The parameter $B_0$ is fixed to be 0.025 throughout the study, and the corresponding wavelength of the fastest growing mode (FGM) of the MRI is $\lambda_{\rm FGM} = 2 \pi v_{A}/\Omega = 0.16$ where $v_{A}=B_0/\sqrt{\rho}$ denotes the {\Alfven} velocity.
Following \cite{2004ApJ...605..321S}, we investigate the gas pressure dependence of the saturation level by varying the initial plasma beta value as $10^2,10^3,10^4,10^5$, and $10^6$.
The simulation is relevant to actual accretion disks when the vertical scale size $L$ is comparable to the pressure scale height $\sqrt{2P/\rho}/\Omega$ at $\beta = 10^4$.
The simulation runs more than a 100 orbit periods to improve the statistics.
We measure the statistical variables by averaging in whole space and time over the last 50 orbit periods.

\section{RESULTS}\label{sec:results}
\subsection{Ideal Magnetohydrodynamic Simulations}\label{sec:norm-resol-case}
Tables \ref{tab:sim_table_01} to \ref{tab:sim_table_03} list the simulation parameters and results obtained for the ideal MHD simulation.
As can be noted from the tables, we use different Riemann solvers for the fluid and magnetic field updates.
Table \ref{tab:sim_table_01} lists results for the application of the HLLR Riemann solver for the fluid and the HLL Riemann solver for the magnetic field.
Table \ref{tab:sim_table_02} lists the results for the case when the HLL Riemann solver is used for both the fluid and magnetic field.
Table \ref{tab:sim_table_03} corresponds to the case when the HLLR Riemann solver is used for both the fluid and magnetic field.
Column (1) specifies the model applied.
The first letter I denotes ``ideal''.
The second letter denotes the 
{logarithmic} of the plasma beta value (listed in column (4)).
The remaining letters represent abbreviation of the reconstruction function and the Riemann solver (W = WENO, Z = WZMP, H = HLL, R = HLLR).
Columns (2) and (3) specify the reconstruction function used for the fluid and magnetic field, respectively.
Columns (4) and (5) list the initial plasma beta and gas pressure, respectively.
Columns (6) and (7) list the statistical average of the magnetic energy $E_{M} = \vect{B}^2/8 \pi$ and the kinetic energy $E_{K} = \rho \vect{u}'^2/2$, respectively.
Columns (8) and (9) list the statistical average of the Maxwell stress $w_{M} = -B_x B_y/4 \pi$ and the Reynolds stress $w_{R} = \rho u_x' u_y'$, respectively.
Column (10) lists the $\alpha$ parameter given by \cite{1973A&A....24..337S}, $\alpha = (w_{M}+w_{R})/P_0$.
Finally, column (11) lists the ``numerical'' magnetic Prandtl number, which is estimated in a later section.

\begin{figure}[p]
\centering
\epsscale{1.0}
\plotone{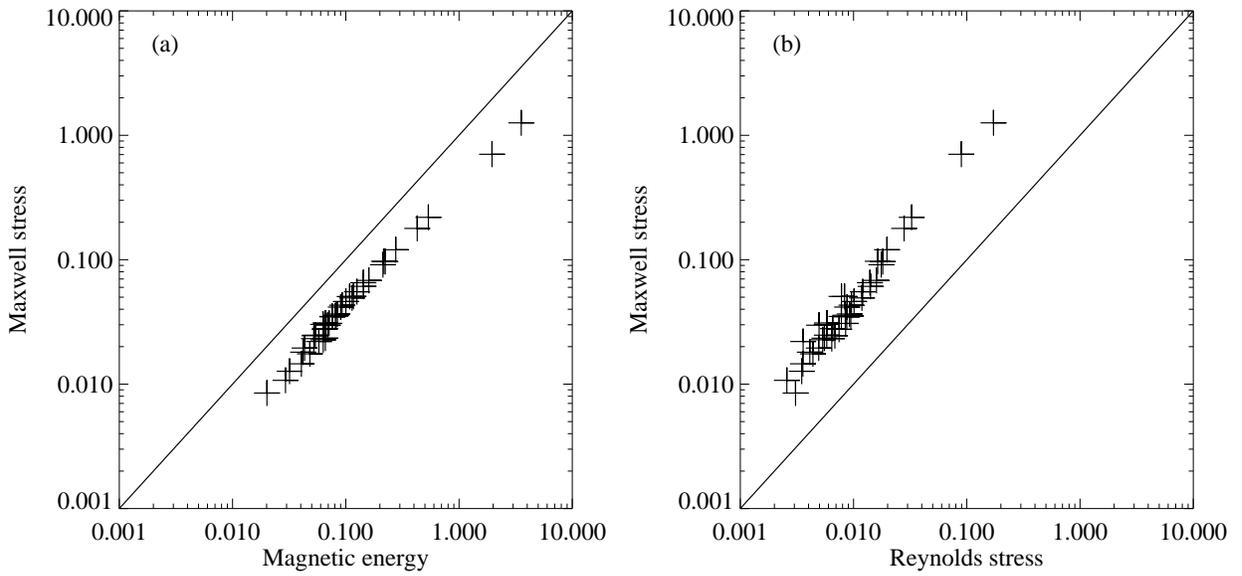}
\caption{Correlation of statistical averages in ideal MHD simulations of the MRI-induced turbulence between (a) the magnetic energy and the Maxwell stress, and (b) the Reynolds stress and the Maxwell stress.}
\label{fig:bsqr-bstrs}
\end{figure}

\begin{figure}[p]
\centering
\epsscale{1.0}
\plotone{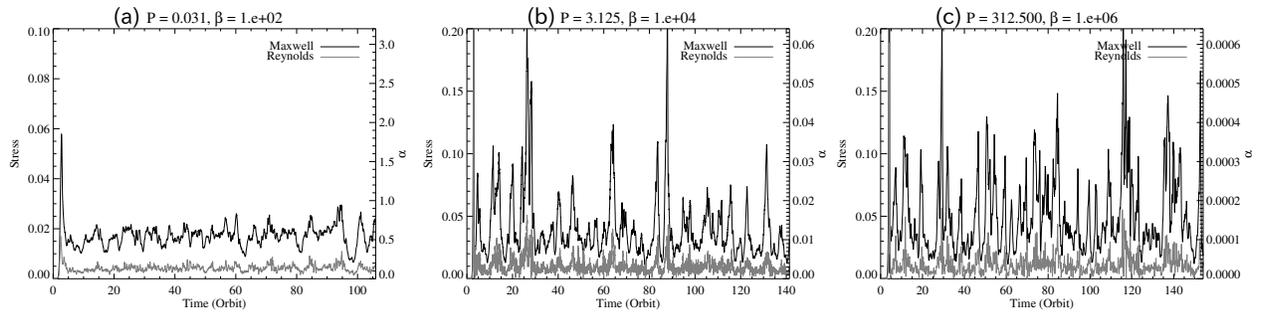}
\caption{Time profile of the Maxwell (black line) and Reynolds (gray line) stresses in the runs (a) I2-WRZH, (b) I4-WRZH, and (c) I6-WRZH.}
\label{fig:strs-n4-wrzh}
\end{figure}

\begin{figure}[p]
\centering
\epsscale{1.0}
\plotone{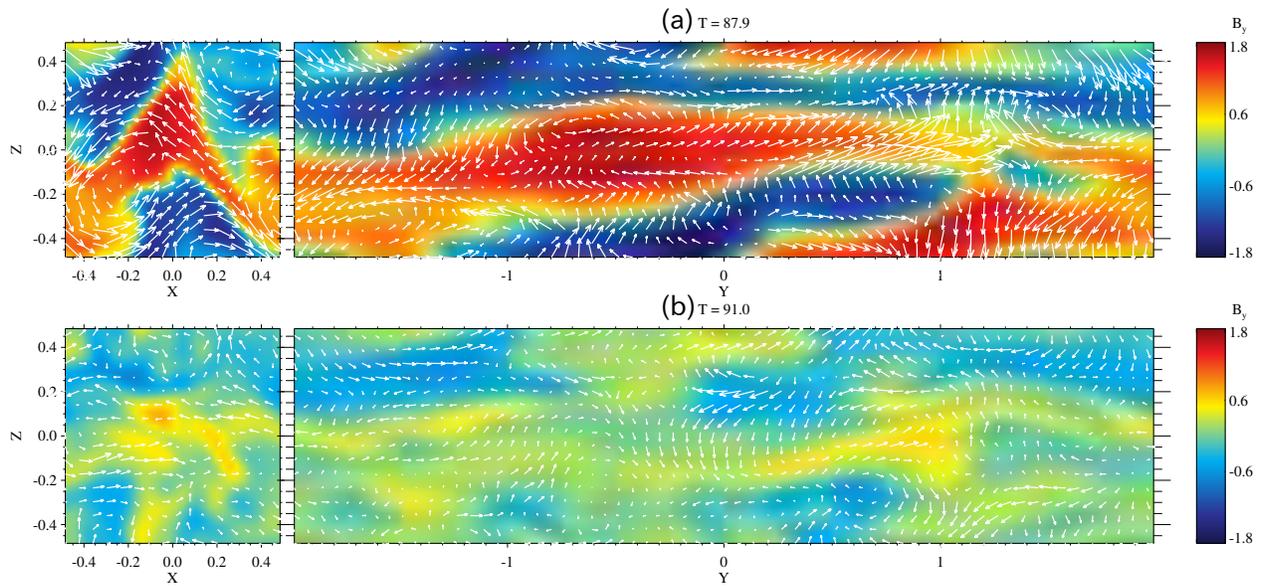}
\caption{Spatial profile of the $y$-component of the magnetic field at (a) peak and (b) decay periods in the run I4-WRZH. Left panels show the $x-z$ plane at $y=-2$, and right panels show the $y-z$ plane at $x=0.5$. White arrows indicate the direction of flow (background shear is subtracted).}
\label{fig:ux_in_xz}
\end{figure}

First, we verify the correlation of statistical averages in the turbulent state (Figure \ref{fig:bsqr-bstrs}). 
All the simulations listed in Tables \ref{tab:sim_table_01}-\ref{tab:sim_table_03} are considered.
The Maxwell stress is smaller than the magnetic energy by a factor of $\sim 2$ and larger than the Reynolds stress by a factor of $\sim 5$ within the range of two orders of magnitude.
The Maxwell stress, magnetic energy, and the Reynolds stress show good correlation, and the results appear consistent with those of previous studies \citep[e.g.,][]{1995ApJ...440..742H,1996ApJ...464..690H,2004ApJ...605..321S}.
{ Figure \ref{fig:strs-n4-wrzh} shows time profiles of the Maxwell and Reynolds stresses in three runs denoted as (a) I2-WRZH ($P_0 = 0.03125$), (b) I4-WRZH ($P_0 = 3.125$), and (c) I6-WRZH ($P_0 = 312.5$).
Many random spikes are prominent for both the Maxwell and Reynolds stresses in the turbulent state for the runs I4-WRZH and I6-WRZH.
Their amplitude is comparable to or larger than the temporally-averaged values.
When the pressure is low (I2-WRZH), on the other hand, the temporally-averaged value is smaller than that in the high pressure cases and spikes are less prominent.
}
Figure \ref{fig:ux_in_xz} shows the $y$-component of the magnetic field at the (a) peak and (b) decay of a spike in the run I4-WRZH.
Large-scale flows are well developed at the peak.
The amplitude of the flows is comparable to the background shear flow.
The magnetic field is stretched and amplified predominantly along the $y$-direction.
{ The spatially-averaged magnetic energy is increased by a factor of several hundred from the initial value, thereby giving plasma beta of several tens.
Consequently, the flow velocity that is comparable to the {\Alfven} speed is still much slower than the sound speed.
}
At the decay, the flow collapses to a smaller scale due to secondary parasitic instabilities \citep{1994ApJ...432..213G}.
The growth of the large-scale flow and its subsequent decay repeatedly appear in the turbulent state.
These features are in good agreement with the results obtained by \cite{2001ApJ...561L.179S}.
Therefore, we can conclude that the simulation successfully solves the MHD turbulence induced by the MRI.

\begin{figure}[p]
\centering
\epsscale{1.0}
\plotone{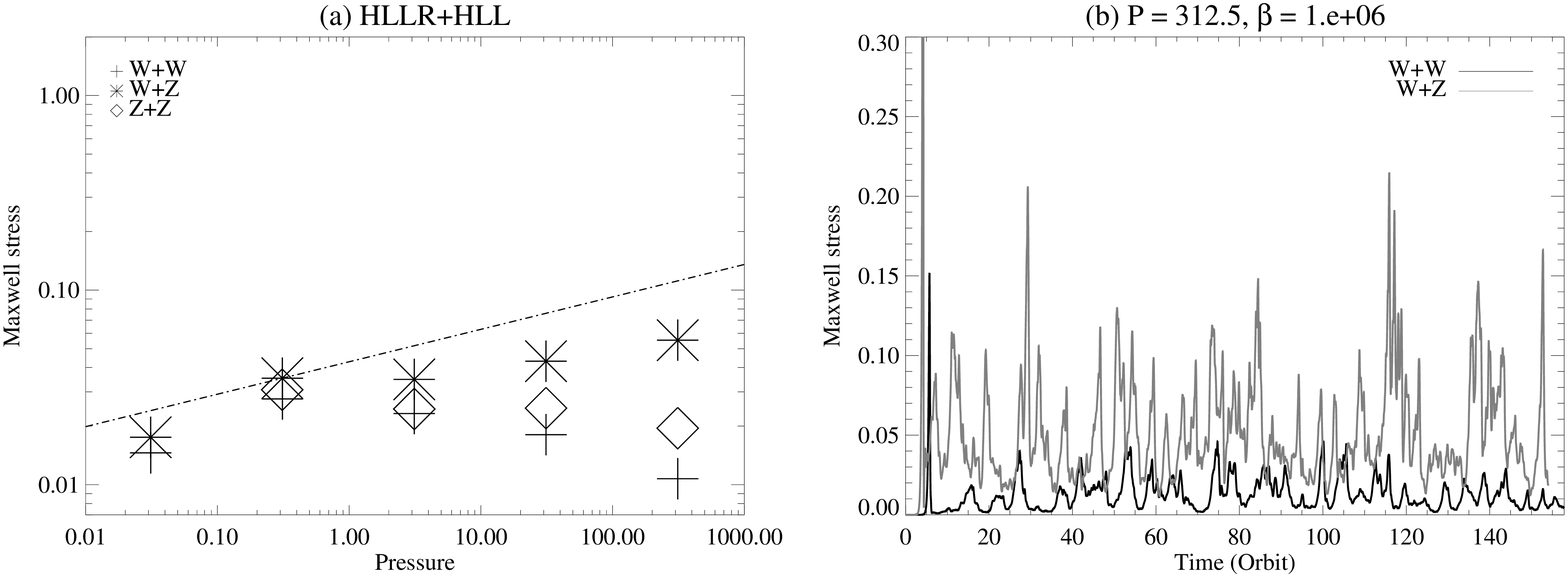}
\caption{(a) Statistical average of the Maxwell stress as a function of the gas pressure. The HLLR and HLL Riemann solvers are used for the fluid and magnetic field updates, respectively. The various symbols represent simulation results obtained with different reconstruction functions. The dash-dotted line represents the correlation $w_{M} \propto P^{1/6}$ as obtained by \cite{2004ApJ...605..321S}. (b) Time profile of the Maxwell stress in the runs I6-WRWH (black line) and I6-WRZH (gray line), corresponding the cross and asterisk, respectively, at $P=312.5$ in panel (a).}
\label{fig:strs-hllr-hll}
\end{figure}

Following \cite{2004ApJ...605..321S}, we investigate the gas pressure dependence of the saturation level of the Maxwell stress in Figure \ref{fig:strs-hllr-hll}(a).
The data are taken from Table \ref{tab:sim_table_01}, that is, we use the HLLR Riemann solver for the fluid and the HLL Riemann solver for the magnetic field.
There are three combinations of the reconstruction function for the fluid and magnetic field.
The crosses correspond to the results obtained with the WENO reconstruction for both the fluid and magnetic field (I?-WRWH)\footnote{The symbol ``?'' expresses an arbitrary digit from 2 to 6.}.
The asterisks represent the results of the WENO reconstruction for the fluid and the WZMP reconstruction for the magnetic field (I?-WRZH).
This model improves the accuracy of the magnetic field with respect to the first I?-WRWH model.
The diamonds denote the results obtained with the WZMP reconstruction for both the fluid and magnetic field (I?-ZRZH).
This model improves the accuracy of the fluid as well as the magnetic field with respect to the first I?-WRWH model.

\cite{2004ApJ...605..321S} have shown that the saturation level of the Maxwell stress weakly depends on the gas pressure as $w_{M} \propto P^{1/6}$ for the isothermal case.
However, the pressure dependence in our simulation is considerably different from their result.
In the low-pressure range of $P = 0.01-1.0$, the Maxwell stress increases with increasing the pressure regardless of the choice of models.
With increase in the range of pressure $P = 1.0-1000$, the pressure dependence differs among the models.
In the I?-WRWH model, the Maxwell stress is a decreasing function of the pressure.
In the I?-WRZH model, the Maxwell stress is a weakly increasing function of the pressure.
In the I?-ZRZH model, the Maxwell stress is independent of or weakly decreases with the pressure.
The difference is more significant at higher pressures.
Figure \ref{fig:strs-hllr-hll}(b) shows the time profile of the Maxwell stress at the highest pressure with the two models I6-WRWH and I6-WRZH.
There are significant differences in not only the average but also the peak and fluctuation amplitude between the models.
This result indicates that our simulation is very sensitive to the reconstruction function, and hence, 
it fails to evaluate the gas pressure dependence.

\begin{figure}[p]
\centering
\epsscale{1.0}
\plotone{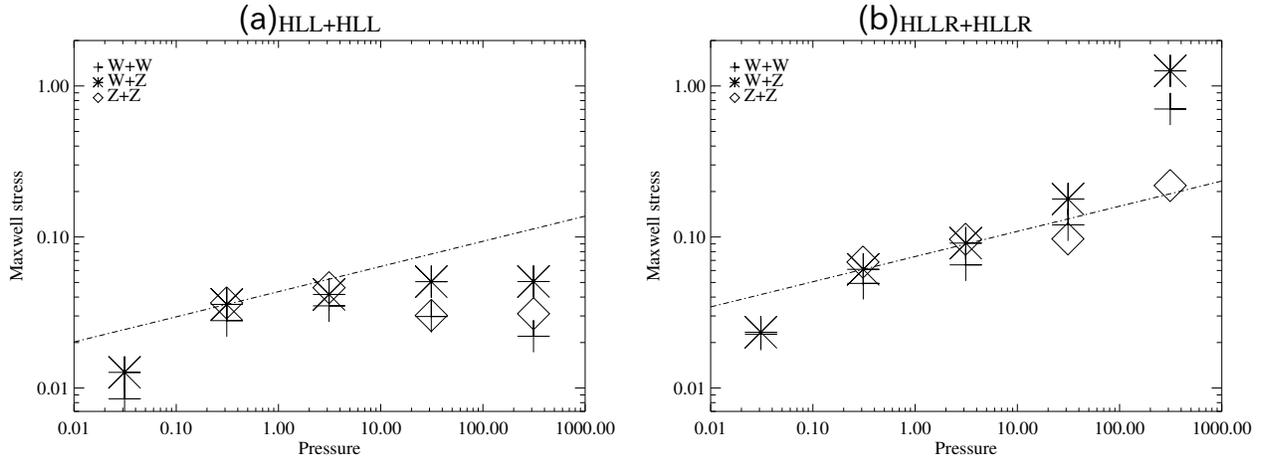}
\caption{Statistical average of the Maxwell stress as a function of the gas pressure. The (a) HLL and (b) HLLR Riemann solver are used. The various symbols represent simulation results obtained with different reconstruction functions. The dash-dotted line represents the correlation $w_{M} \propto P^{1/6}$ as obtained by \cite{2004ApJ...605..321S}.}
\label{fig:strs-hll-hll-hllr-hllr}
\end{figure}

We also examine the effect of the difference of the the Riemann solver on the gas pressure dependence of the saturation level.
In the previous case, there is a mismatch of the Riemann solver between the fluid (HLLR scheme) and the magnetic field (HLL scheme).
Here, we use the same Riemann solver for both the fluid and magnetic field.
We first use the HLL scheme, which degrades the accuracy of the fluid with respect to that in the previous case.
Secondly, we use the HLLR scheme, which improves the accuracy of the magnetic field from the previous case.
The results are listed in Tables \ref{tab:sim_table_02} and \ref{tab:sim_table_03}, and the pressure dependence of the saturation level of the Maxwell stress is shown in Figure \ref{fig:strs-hll-hll-hllr-hllr}.
These results appear to be less sensitive to the reconstruction function than in the previous case, thereby indicating that the use of the same Riemann solver for both the fluid and magnetic field improves the convergence of the simulation.
However, the results are found to be very sensitive to the Riemann solver itself.
The Maxwell stress is almost independent of the pressure in the high-pressure range of $P = 1.0-1000$ with the F1-HLL-F2-HLL scheme (Figure \ref{fig:strs-hll-hll-hllr-hllr}(a)).
The amplitude is comparable to or slightly larger than that obtained with the F1-HLLR-F2-HLL scheme. 
On the other hand, the Maxwell stress is an increasing function of the pressure within the explored parameter range with the F1-HLLR-F2-HLLR scheme (Figure \ref{fig:strs-hll-hll-hllr-hllr}(b)).
The amplitude is considerably larger than those in the previous cases with the F1-HLLR-F2-HLL and F1-HLL-F2-HLL schemes, particularly at higher pressures.
In the intermediate pressure range of $P = 0.1-100$, the result follows the correlation by \cite{2004ApJ...605..321S} (dash-dotted line).

The effects of numerical {dissipation} on the gas pressure dependence of the saturation level are briefly summarized as follows:
(i) When the {accuracy} of the magnetic field is improved, the saturation level increases. (ii) When the {accuracy} of the fluid is improved, the saturation level decreases.
This trend indicates that the saturation level depends on the numerical magnetic Prandtl number, $Pr_m = \nu /\eta$, where $\nu,\eta$ are the kinematic viscosity and resistivity, respectively.

\subsubsection{Numerical Magnetic Prandtl Number}\label{sec:num-magn-prandtl}
The magnetic Prandtl number dependence of the MRI-induced turbulence was first reported by \cite{2007MNRAS.378.1471L} for finite net magnetic flux cases ($B_z = {\rm const}$), and it has been studied by \cite{2007A&A...476.1123F} for zero net flux cases ($B_z \propto \sin(2 \pi x)$).
They have carried out three-dimensional visco-resistive MHD simulations under the shearing-box approximation, and they have shown that the saturation level of the Maxwell stress is an increasing function of the magnetic Prandtl number.
\cite{2007MNRAS.378.1471L} have shown that $w_{M} \propto Pr_m^{0.25-0.5}$ for the explored range of $0.12 < Pr_m < 8.0$.
\cite{2007A&A...476.1123F} have determined the critical magnetic Prandtl number as $Pr_{m,{\rm crit}} \sim 1$ for zero net flux cases, below which turbulence is quenched.

{ Our ideal MHD simulations could not evaluate the gas pressure dependence of the saturation level of the Maxwell stress. It depends on the choice of the numerical scheme.} 
Based on the studies by \cite{2007MNRAS.378.1471L} and \cite{2007A&A...476.1123F}, we speculate that
 the numerical magnetic Prandtl number depends on the choice of the numerical scheme and varies with the pressure.
To verify it, let us consider the propagation of incompressible {\Alfven} waves along the $x$-direction,
\begin{eqnarray}
\pdif{\vect{v}_{t}}{t} &=& \frac{B_x}{4 \pi \rho} \pdif{\vect{B}_t}{x},\label{eq:6}\\
\pdif{\vect{B}_{t}}{t} &=& B_x \pdif{\vect{v}_t}{x},\label{eq:7}
\end{eqnarray}
where $\vect{v}_t$ and $\vect{B}_t$ denote the transverse components of the velocity and magnetic field, $\rho$ and $B_x>0$ are uniform, and $v_x = 0$ is assumed.
As is well known, the combination of these equations gives the advection equation for the \Elsasser variables $\vect{f}^{\pm}=\vect{v}_t \pm \vect{v}_{At}$,
\begin{eqnarray}
\pdif{\vect{f}^{\pm}}{t}  \mp v_{Ax} \pdif{\vect{f}^{\pm}}{x} = 0.\label{eq:8}
\end{eqnarray}

To evaluate the numerical error in the {\Alfven} wave propagation problem, we discretize the equations (\ref{eq:6}) and (\ref{eq:7}) in space by means of a first-order local Lax$-$Friedrichs scheme (equivalent to the HLL scheme in the homogeneous case),
\begin{eqnarray}
\dif{\vect{v}_{t,i}}{t} &=& \frac{v_{Ax}}{\sqrt{4 \pi \rho}} \frac{\vect{B}_{t,i+1}-\vect{B}_{t,i-1}}{2 \Delta} + \frac{c}{2 \Delta} \left(\vect{v}_{t,i+1}-2\vect{v}_{t,i}+\vect{v}_{t,i-1}\right) + O\left(\Delta\right),\label{eq:9}\\
\dif{\vect{B}_{t,i}}{t} &=& B_x \frac{\vect{v}_{t,i+1}-\vect{v}_{t,i-1}}{2 \Delta} + \frac{c}{2 \Delta} \left(\vect{B}_{t,i+1}-2\vect{B}_{t,i}+\vect{B}_{t,i-1}\right) + O\left(\Delta\right),\label{eq:10}
\end{eqnarray}
where $c$ denotes the maximum speed among eigenmodes. These equations yield the advection equation for the \Elsasser variables in a discrete form as
\begin{eqnarray}
\dif{\vect{f}_{i}^{+}}{t} - v_{Ax} \frac{\vect{f}_{i+1}^{+}-\vect{f}_{i}^{+}}{\Delta} &=& \frac{|c-v_{Ax}|\Delta}{2} \frac{\vect{f}_{i+1}^{+}-2\vect{f}_{i}^{+}+\vect{f}_{i-1}^{+}}{\Delta^2} + O\left(\Delta\right),\label{eq:11}\\
\dif{\vect{f}_{i}^{-}}{t} + v_{Ax} \frac{\vect{f}_{i}^{-}-\vect{f}_{i-1}^{-}}{\Delta} &=& \frac{|c-v_{Ax}|\Delta}{2} \frac{\vect{f}_{i+1}^{-}-2\vect{f}_{i}^{-}+\vect{f}_{i-1}^{-}}{\Delta^2} + O\left(\Delta\right).\label{eq:12}
\end{eqnarray}
The left-hand-side corresponds to the first-order upwind discretization.
In addition to the discretization error $O(\Delta)$, there is a second-order diffusion term on the right-hand-side, which acts as the resistivity to dissipate {\Alfven} waves.
The coefficient of this numerical diffusion, $\eta_{\rm num} = |c-v_{Ax}|\Delta/2$, is negligible for low beta $(c = c_{\rm fast} \sim v_{Ax})$, but not for high beta, $\eta_{\rm num} \propto c_{\rm fast} \propto P^{1/2}$.
The use of single-state Riemann solvers such as the Lax$-$Friedrichs and HLL schemes leads to the pressure-dependent numerical diffusion of {\Alfven} waves.
{ Multi-state Riemann solvers such as the Roe \citep{1981JCoPh..43..357R}, HLLR, and HLLD \citep{2005JCoPh.208..315M} schemes significantly reduce the amount of numerical dissipation} because they are designed to resolve the {\Alfven} mode as well as the fast mode.

However, the above statement holds only for one-dimensional simulations. In actual multi-dimensional simulations, even multi-state Riemann solvers lead to the pressure-dependent numerical diffusion of {\Alfven} waves owing to an error associated with the directional splitting (the one-dimensional Riemann solver is applied dimension-by-dimension).
Numerical diffusion of MHD turbulence is inevitable, and it depends on the plasma beta.
Therefore, we consider that the pressure dependence of the saturation level of the MRI-induced turbulence is indistinguishable from the numerical magnetic Prandtl number dependence.

To quantitatively discuss the effect of the numerical magnetic Prandtl number on the saturation level of the Maxwell stress, we estimate numerical viscosity and resistivity as follows.
We restart simulations with $\Omega=0$ to numerically dissipate turbulence.
Subsequently, the spatially averaged decay rates are measured as the numerical viscosity $\nu_{\rm num}$ and resistivity $\eta_{\rm num}$ via the following equations obtained by integrating viscous momentum equation and resistive induction equation over space,
\begin{eqnarray}
\nu_{\rm num} &=& \left[-\pdif{}{t}\overline{\frac{\vect{u}'^2}{2}} + \overline{\left\{\frac{\vect{u}' \cdot \left(\vect{j} \times \vect{B}\right)}{\rho}\right\}}\right] / \left[ \overline{\left(\nabla \times \vect{u}' \right)^2} + \left(4/3\right) \overline{\left(\nabla \cdot \vect{u}' \right)^2}\right], \label{eq:13}\\
\eta_{\rm num} &=& \left[-\pdif{}{t}\overline{\frac{\vect{B}^2}{8 \pi}} - \overline{\vect{u}' \cdot \left(\vect{j} \times \vect{B}\right)}\right] / \overline{\vect{j}^2},\label{eq:14}
\end{eqnarray}
where $\vect{j} = \nabla \times \vect{B}/4 \pi$ denotes the current density, and the overline represents spatial averaging.
{The first term in the numerator represents the energy change, and the second term represents the work done by the dynamo.}
The spatial variation in the density is ignored for simplicity.
{Note that this method is inaccurate for strongly compressible cases.}
Finally, we estimate the numerical magnetic Prandtl number as $Pr_{m,{\rm num}} = \nu_{\rm num}/\eta_{\rm num}$.
This is a crude estimation.
In reality, numerical diffusion coefficients depend on the wavenumber, thus depend on the period over which the MHD simulation data are taken.
\cite{2007A&A...476.1113F} have proposed a more rigorous treatment to estimate the numerical resistivity by analyzing the induction equation in Fourier space.
We adopt the above method for simplicity, and we assume that the numerical viscosity and resistivity exhibit a similar dependence on the wavenumber, and consequently, the numerical magnetic Prandtl number is insensitive to the wavenumber. 

\begin{figure}[p]
\centering
\epsscale{1.0}
\plotone{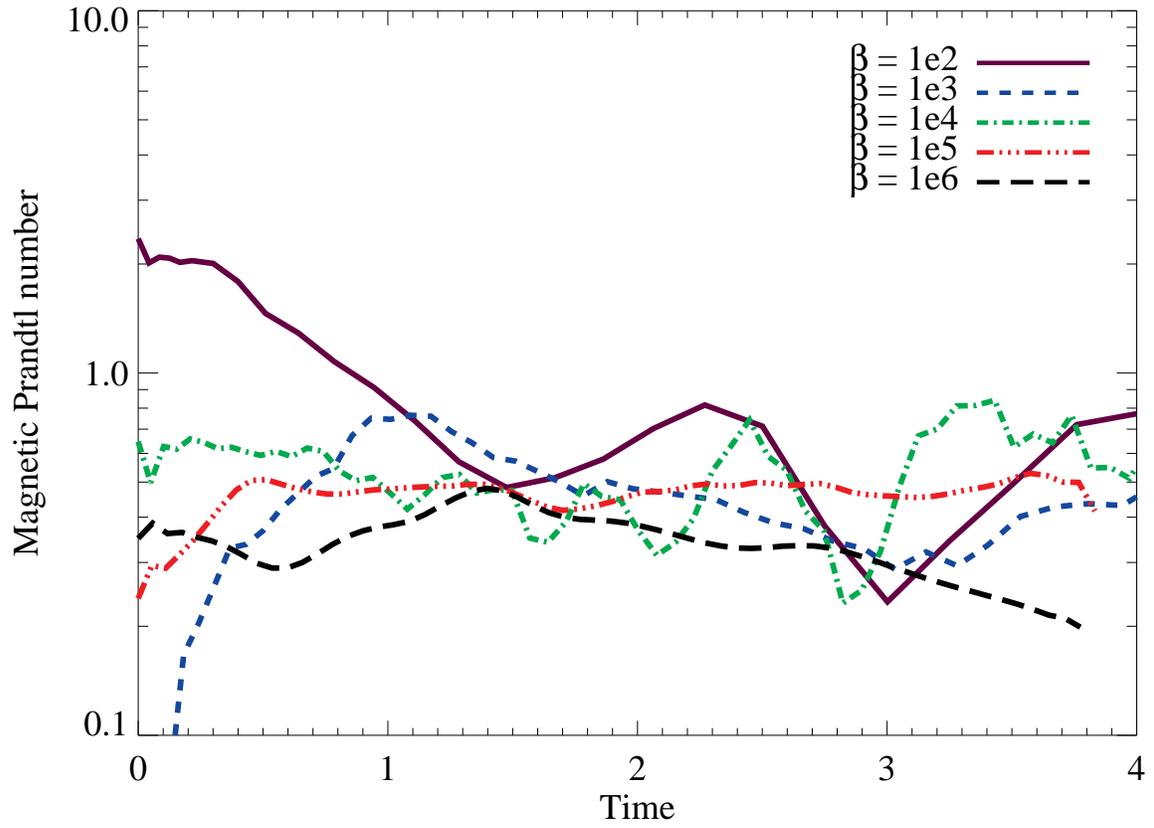}
\caption{Example of time profile of the numerical magnetic Prandtl number. The WENO-HLLR-WENO-HLL scheme is used. The various lines represent simulation results obtained with different initial plasma beta values.}
\label{fig:mpran_140130}
\end{figure}

Example estimations of the numerical magnetic Prandtl number are shown in Figure \ref{fig:mpran_140130} for the WENO-HLLR-WENO-HLL scheme.
In this scheme, the HLL Riemann solver is used only for the magnetic field, which degrades the {accuracy} of the magnetic field when compared with that of the fluid as the pressure increases.
Therefore, the numerical magnetic Prandtl number is expected to be a decreasing function of the pressure.
This is confirmed in our estimation;
the temporally averaged values are $0.76,0.43,0.51,0.45,0.36$ for $\beta=10^2,10^3,10^4,10^5,10^6$, respectively.

The averaged values of the numerical magnetic Prandtl number in all simulation runs are listed in column (11) of Tables \ref{tab:sim_table_01}-\ref{tab:sim_table_03}.
We observe the following systematic trends:
(i) When both the fluid and magnetic field are updated with the HLL scheme (Table \ref{tab:sim_table_02}), the numerical magnetic Prandtl number is close to unity and almost independent of the pressure.
(ii) When the {accuracy} of the fluid is improved by the HLLR scheme (from Table \ref{tab:sim_table_02} to Table \ref{tab:sim_table_01}), the numerical magnetic Prandtl number clearly decreases and is a decreasing function of the pressure. 
(iii) When the {accuracy} of the magnetic field is improved by the HLLR scheme (from Table \ref{tab:sim_table_01} to Table \ref{tab:sim_table_03}), the numerical magnetic Prandtl number clearly increases.
(iv) When both the fluid and magnetic field are updated with the HLLR scheme (Table \ref{tab:sim_table_03}), the numerical magnetic Prandtl number is larger than unity, and it is an increasing function of the pressure.
(v) When the WZMP reconstruction function is utilized to improve the {accuracy} of the magnetic field, the numerical magnetic Prandtl number clearly increases.
(vi) When the WZMP reconstruction function is utilized to improve the {accuracy} of the fluid, the numerical magnetic Prandtl number clearly decreases.
The numerical magnetic Prandtl number is smaller or larger than unity, and its pressure dependence is not straightforward, instead depending on the choice of the Riemann solver as well as the reconstruction function.

\begin{figure}[p]
\centering
\epsscale{1.0}
\plotone{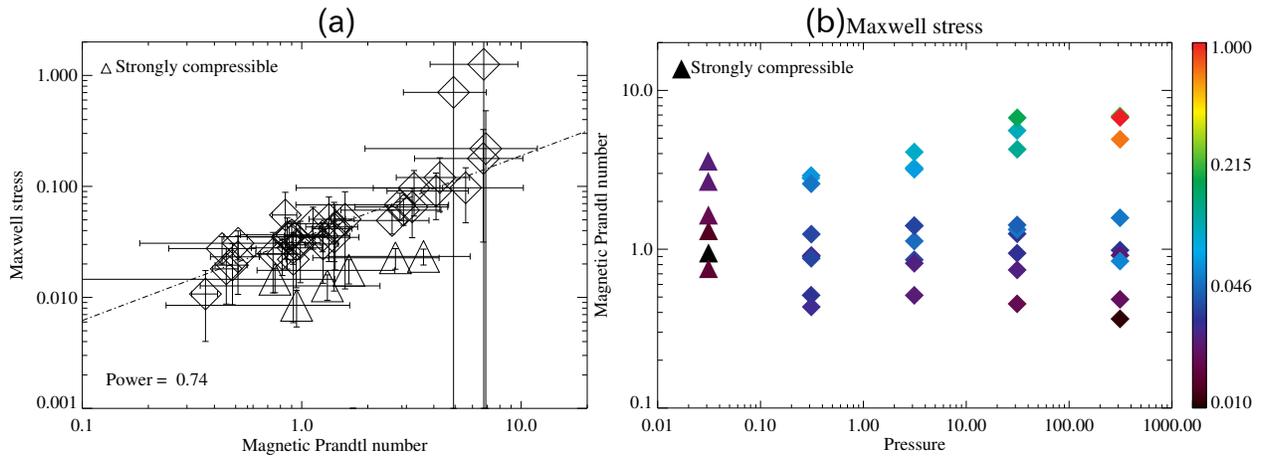}
\caption{Statistical average of the Maxwell stress (a) as a function of the numerical magnetic Prandtl number, and (b) as a function of the gas pressure and the numerical magnetic Prandtl number}
\label{fig:mpran_strs}
\end{figure}

In Figure \ref{fig:mpran_strs}(a), the Maxwell stress is shown as a function of the numerical magnetic Prandtl number.
Error bars show the standard deviation of the temporal average.
Positive correlation is clearly seen between the two parameters.
If we exclude the data in strongly compressible cases ($\beta=10^2$, triangles) and the data with extremely large errors, we obtain the correlation $w_{M} \propto Pr_{m,{\rm num}}^{0.74}$ (dash-dotted line).
In Figure \ref{fig:mpran_strs}(b), the Maxwell stress is shown as a function of the gas pressure and the numerical magnetic Prandtl number.
Except for strongly compressible cases, the Maxwell stress is more sensitive to the numerical magnetic Prandtl number than the pressure.
{ Solutions of the ideal MHD simulation of the MRI-induced turbulence are found to be subject to the numerical magnetic Prandtl number of the applied scheme.}
Therefore, we conclude that it is necessary for the MRI-induced turbulence simulation to use explicit viscosity and resistivity to control the magnetic Prandtl number.
The visco-resistive MHD simulation is indispensable to discuss the property of the MRI-induced turbulence.

{ 
\subsubsection{Convergence}\label{sec:convergence}
In ideal MHD simulations, numerical dissipation scales are related to numerical resolution (number of grid points) as well as characteristics of the applied scheme.
Previous studies have addressed the numerical convergence of solutions of the MRI under an unstratified local shearing box approximation, and they have found that it depends on the initial magnetic field configuration; increasing the resolution decreases the saturation level for zero net flux cases \citep{2007A&A...476.1123F,2007ApJ...668L..51P,2011ApJ...739...82B} whereas it increases the level for finite net flux cases\footnote{This will hold at low and moderate resolutions. Since it cannot grow indefinitely, the level should be converged to an asymptotic value for high enough resolution.} \citep{2008A&A...487....1B,2008MNRAS.385.1036S}.

To address the convergence of our simulation results, we conduct the same runs with doubling the resolution (grid points of $64\times256\times64$).
We use the three combinations of the Riemann solver that are same as the previous low resolution runs, and the WENO reconstruction is adopted for both the fluid and magnetic field throughout this subsection.
Table \ref{tab:sim_table_highideal} lists simulation parameters and results. The format is same as that used in Tables \ref{tab:sim_table_01}-\ref{tab:sim_table_03} except that the first letter H in column (1) denotes ``high resolution'' and columns (2) and (3) specify the Riemann solver used for the fluid and magnetic field, respectively.

Figure \ref{fig:strs_highres_ideal} compares the Maxwell stress in the low and high resolution runs.
Except for the case at $P=312.5$ with the WENO-HLLR-WENO-HLLR scheme, the Maxwell stresses in the high resolution runs are larger than those in the low resolution runs.
\cite{2009ApJ...694.1010G} have argued that the increase in the saturation level is attributed to the resolution of small-scale structures near the correlation length at which energy is injected.
We do not achieve the convergence within currently available computational resources.
Meanwhile, we find that the numerical magnetic Prandtl number dependence are still observed in the high resolution runs; the Maxwell stress is low in the H?-WRWH model, medium in the H?-WHWH model, and high in the H?-WRWR model.
Using the method in Section \ref{sec:num-magn-prandtl}, the numerical magnetic Prandtl numbers are estimated to be $Pr_{m,{\rm num}}\sim 0.5,1.0,{\rm and} \; 2.0$ in the H?-WRWH, H?-WHWH, and H?-WRWR models, respectively.
The Maxwell stress roughly obeys the expression $w_{M} \propto Pr_{m,{\rm num}}^{0.5-1.0}$ which is similar to that obtained in the low resolution runs.
Note that the numerical magnetic Prandtl number is the ratio of the numerical viscosity to resistivity which are inversely proportional to the resolution of fluid and magnetic field, respectively.
If the numerical scheme is designed so that the orders of accuracy of fluid and magnetic field are nearly equal, the numerical magnetic Prandtl number is expected to be insensitive to the number of grid points.
It rather sensitive to characteristics of the applied scheme.
This implies that different ideal MHD simulation codes may lead to different results depending on the numerical magnetic Prandtl number,  even if sufficient grid points are used to confirm numerical convergence.
}

\subsection{Visco-resistive Magnetohydrodynamic Simulations}\label{sec:high-resolution-with}
We perform visco-resistive MHD simulations of the MRI to examine the gas pressure dependence of the saturation level at a constant magnetic Prandtl number.
The viscous and resistive terms are advanced by an operator-splitting method to primitive variables,
\begin{eqnarray}
\pdif{\vect{u}'}{t} &=& \nu \left[ - \nabla \times \nabla \times \vect{u}' + \frac{4}{3} \nabla \left(\nabla \cdot \vect{u}'\right)\right],\label{eq:15}\\
\pdif{\vect{B}}{t} &=& -\eta \nabla \times \nabla \times \vect{B},\label{eq:16}
\end{eqnarray}
where uniform viscosity and resistivity are assumed. 
A simple second-order central difference method is used to discretize these equations.
The viscosity and resistivity vary as $(0.5,1,2,4) \times 10^{-3}$, and the ``physical'' magnetic Prandtl number $Pr_{m} = \nu/\eta $ varies as $0.5, 1, 2, 4$.
The initial plasma beta value varies as $10^3,10^4,10^5,10^6$ (omitting the strongly compressible case of $\beta = 10^2$).
The WENO reconstruction and the HLLR Riemann solver are used for both the fluid and magnetic field throughout the visco-resistive MHD simulation.
The number of grid points is $64\times256\times64$.
The other parameters and numerical techniques are identical to those used in the previous ideal MHD simulation.

Table \ref{tab:sim_table_04} lists simulation parameters and results obtained for the visco-resistive MHD simulation.
The format is similar to that used in Tables \ref{tab:sim_table_01}-\ref{tab:sim_table_03}.
Column (1) specifies the model applied.
The first letter V denotes ``visco-resistive.''
The second to fourth numbers denote the 
{logarithmic} of the plasma beta, the viscosity (multiplied by $10^3$), and the magnetic Prandtl number, respectively.
Columns (2) and (3) list the viscosity and resistivity, respectively.

Figure \ref{fig:bsqr-bstrs-vr} shows the correlation of statistical averages in the turbulent state. 
The Maxwell stress is smaller than the magnetic energy by a factor of $2 - 3$, and larger than the Reynolds stress by a factor of $5 - 10$.
The correlation is consistent with the ideal MHD case (Figure \ref{fig:bsqr-bstrs}).
The magnetic field is anisotropic, and it is preferentially enhanced along the $y$-direction.

\begin{figure}[p]
\centering
\epsscale{1.0}
\plotone{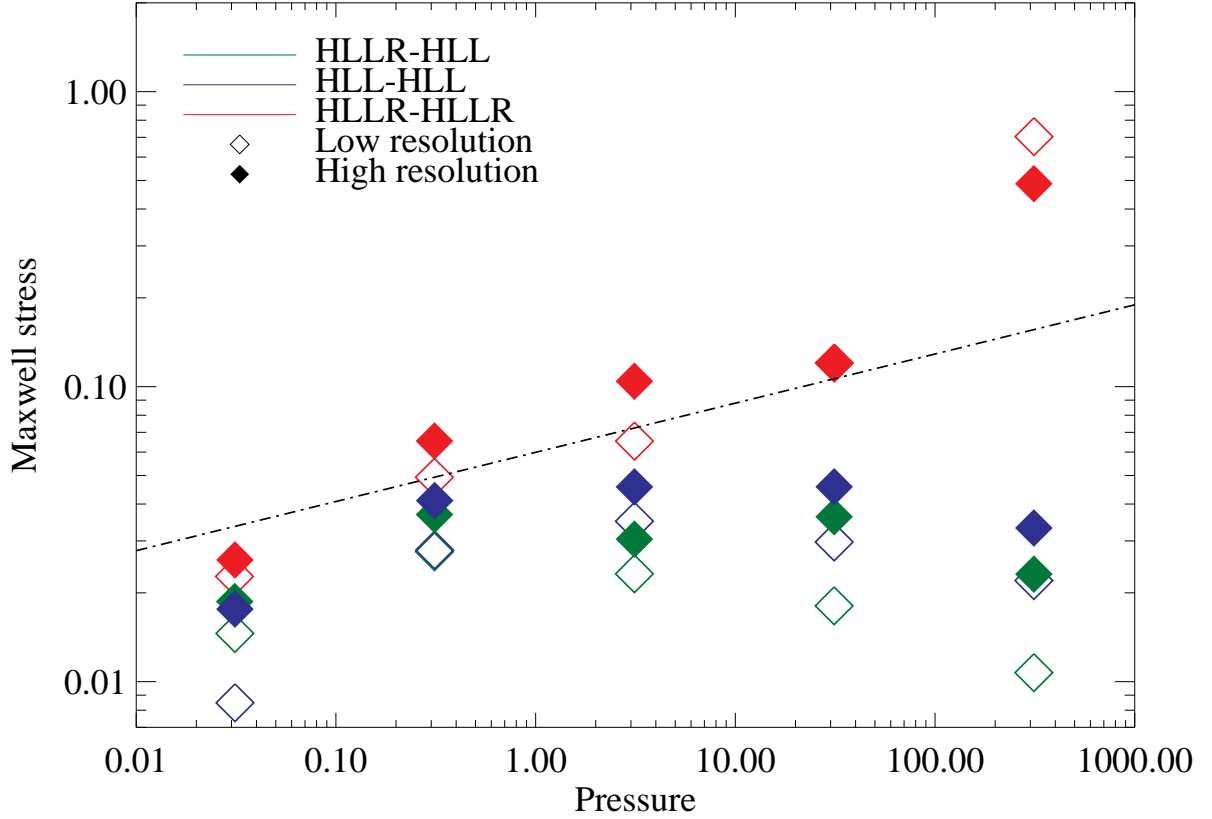}
\caption{Statistical average of the Maxwell stress as a function of the gas pressure. Open and filled diamonds correspond to low and high resolution runs, respectively. The various colors represent simulation results obtained with different combinations of the Riemann solver. The dash-dotted line represents the correlation $w_{M} \propto P^{1/6}$ as obtained by \cite{2004ApJ...605..321S}.
Note that the symbol for the low resolution with the WENO-HLLR-WENO-HLLR scheme (red) is overlapped with that for the high resolution at $P=31.25$.}
\label{fig:strs_highres_ideal}
\end{figure}

\begin{figure}[p]
\centering
\epsscale{1.0}
\plotone{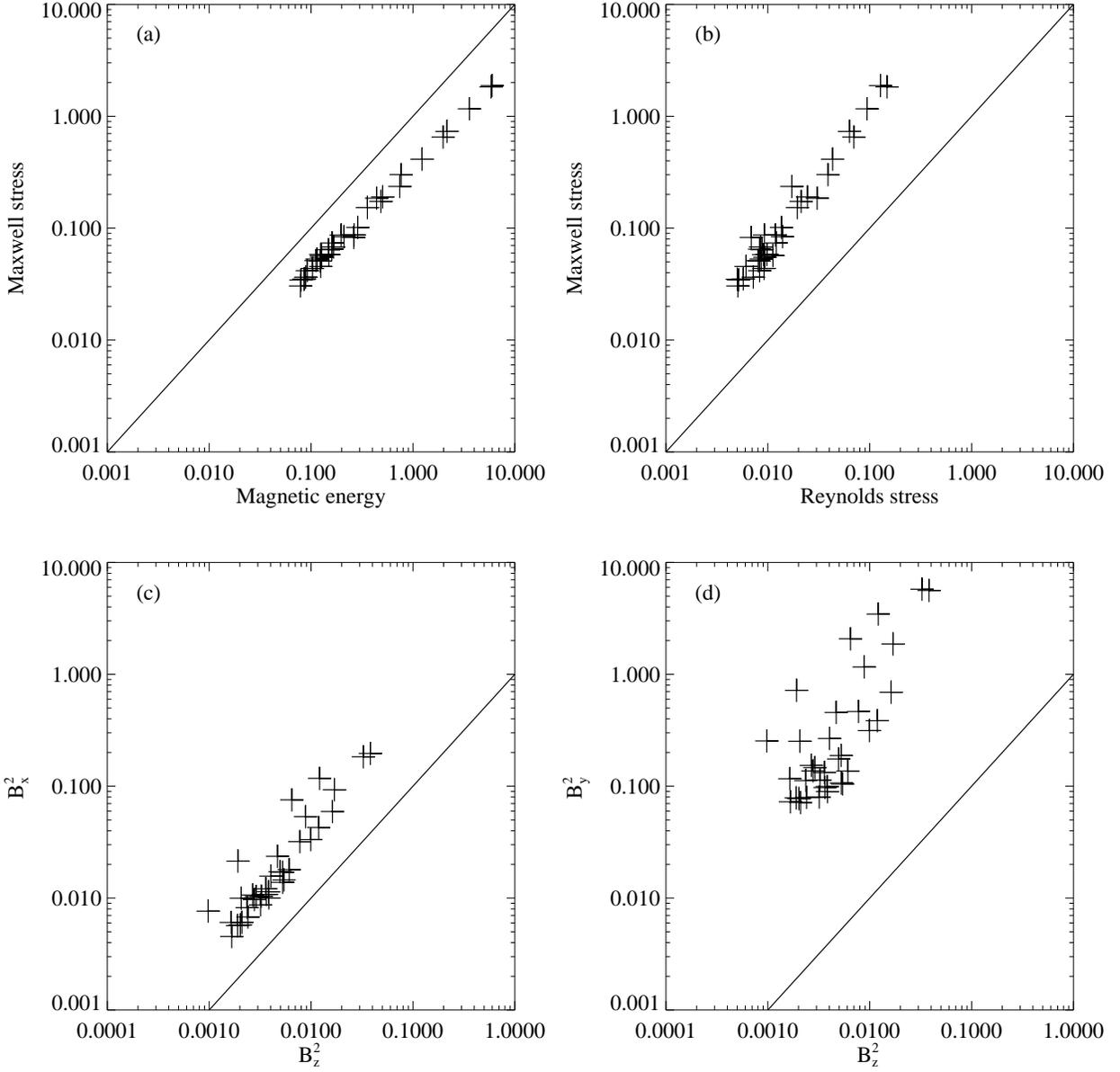}
\caption{Correlation of statistical averages in visco-resistive MHD simulations of the MRI-induced turbulence between (a) the magnetic energy and the Maxwell stress, (b) the Reynolds stress and the Maxwell stress, (c) $B_z^2$ and $B_x^2$, and (d) $B_z^2$ and $B_y^2$.}
\label{fig:bsqr-bstrs-vr}
\end{figure}

\begin{figure}[p]
\epsscale{1.0}
\plotone{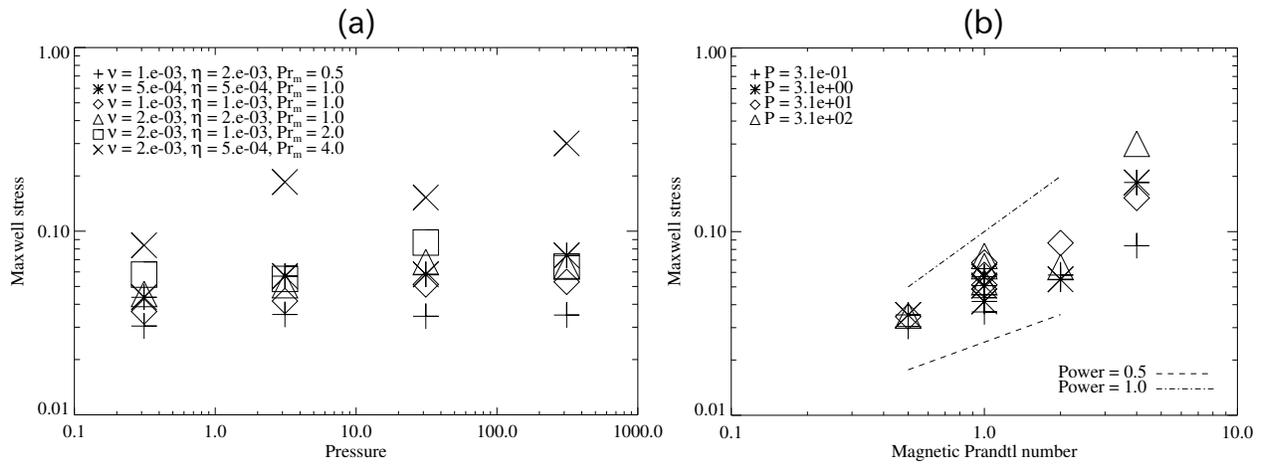}
\caption{Statistical average of the Maxwell stress in visco-resistive MHD simulations as a function of (a) the gas pressure and (b) the physical magnetic Prandtl number. The various symbols represent simulation results obtained with different combinations of the viscosity and resistivity.}
\label{fig:strs_vrmhd}
\end{figure}

Figure \ref{fig:strs_vrmhd} presents the statistical average of the Maxwell stress over six runs: V?-0-1, V?-1-1, V?-2-1, V?-2-2, V?-2-4, and V?-1-0.
In Figure \ref{fig:strs_vrmhd}(a), the Maxwell stress is very weakly dependent on the pressure when the magnetic Prandtl number is fixed.
Note that the ideal MHD simulation with the same numerical scheme shows positive correlation (cross symbols in Figure \ref{fig:strs-hll-hll-hllr-hllr}(b) and red-colored symbols in Figure \ref{fig:strs_highres_ideal}), wherein the numerical magnetic Prandtl number is an increasing function of the pressure. 
We confirm that the visco-resistive MHD simulation with a different Riemann solver yields a consistent result.
A similar tendency is also observed in the case of the ideal MHD simulation with the F1-HLL-F2-HLL scheme (Figure \ref{fig:strs-hll-hll-hllr-hllr}(a) and blue-colored symbols in Figure \ref{fig:strs_highres_ideal}), wherein the numerical magnetic Prandtl number is nearly constant.
Therefore, we conclude that the Maxwell stress in the MRI-induced turbulence at a constant magnetic Prandtl number is almost pressure-independent within the explored parameter range. 
Figure \ref{fig:strs_vrmhd}(b) shows the Maxwell stress as a function of the magnetic Prandtl number.
The stress curve roughly obeys the expression $w_{M} \propto Pr_{m}^{0.5-1.0}$.
The power is close to the correlation obtained in Section \ref{sec:norm-resol-case}.
This consistency indicates that the estimation of our numerical magnetic Prandtl number and hence the conclusion in Section \ref{sec:norm-resol-case} are quantitatively valid.

{ In Table \ref{tab:sim_table_04}, we find an unexpectedly high level for the Maxwell stress especially at the cases with high viscosity (e.g., V6-4-4).
Its origin is discussed in Appendix \ref{sec:trans-betw-lamin}.}

\section{DISCUSSION}\label{sec:discussion}
\subsection{Magnetic Prandtl Number Dependence}\label{sec:magn-prandtl-numb}
{ We discuss the magnetic Prandtl number dependence of the saturation level of the MRI from a physical point of view.
As is shown by \cite{2007MNRAS.378.1471L} and \cite{2008ApJ...684..498P}, the linear growth rate of the axisymmetric incompressible MRI is an increasing function of the Reynolds and magnetic Reynolds numbers, thus not of the magnetic Prandtl number.
Consequently, we consider dissipation by secondary instabilities as a possible candidate.} 
It has been widely recognized that the so-called channel flow of the primary MRI mode becomes unstable against secondary three-dimensional parasitic instabilities \citep{1994ApJ...432..213G}.
The instabilities are related to the Kelvin-Helmholtz (KH) and tearing modes.
{ \cite{2009ApJ...698L..72P} have carried out a linear analysis of the secondary parasitic instability under the configuration of the primary MRI mode and have shown the relative contribution of KH and tearing modes to the parasitic instability.
Although they present a rigorous description for the secondary modes, we simply consider the KH and tearing instabilities independently to derive their Reynolds $(R_e)$ and magnetic Reynolds $(R_m)$ number dependence. Details of the linear analysis of the KH and tearing instabilities are described in  Appendix \ref{sec:line-analys-visco-KH-tear}.}

Figure \ref{fig:kh_linear}(a) shows the linear growth rate of the KH instability.
We assume that the {\Alfven} velocity $v_{A}$ is 1.5 times larger than the half velocity jump and an angle between the shear flow and magnetic field $\theta = 60^{\circ}$. 
{
The linear growth rate of the KH instability is almost independent of the Reynolds and magnetic Reynolds numbers when $R_e, R_m > 10^2$.}
A similar conclusion can be drawn for different combinations of $v_{A}$ and $\theta$.
Therefore, the KH instability does not appear to cause the magnetic Prandtl number dependence.

On the other hand, the tearing instability shows magnetic Prandtl number dependence in Figure \ref{fig:kh_linear}(b,c).
The linear growth rate is a decreasing function of the magnetic Reynolds number (Figure \ref{fig:kh_linear}(b)) and an increasing function of the Reynolds number (Figure \ref{fig:kh_linear}(c)) \citep[cf.,][]{1987PhFl...30.1734P}.
{The {\Alfven} velocity with respect to the dominant magnetic field $B_{y}$ is of the order of $10^{-1}$, thereby yielding $R_e \sim R_m = 10^2$ during the saturation state of the MRI.}
The growth rate can be fitted by $\gamma_{\rm tear} \propto R_{m}^{-1/2}R_{e}^{1/4} = Pr_{m}^{-1/4} R_{m}^{-1/4}$ around $R_e \sim R_m = 10^2$, a decreasing function of the magnetic Prandtl number.

The saturation state is assessed by equating the growth rate of the primary (MRI) and the secondary (tearing) modes \citep{2009ApJ...698L..72P}.
In the course of the MRI, the dominant magnetic field $B_y$ stretched along the $y$-direction will become unstable against the tearing mode.
The condition for the saturation may be
\begin{eqnarray}
\gamma_{\rm tear} \propto \left(\frac{v_{Ay}}{\lambda_{\rm MRI}}\right) Pr_{m}^{-1/4} R_{m}^{-1/4}   \sim \gamma_{\rm MRI}.\label{eq:20}
\end{eqnarray}
Using the relation $\gamma_{\rm MRI}/k_{\rm MRI} = v_{Az} = {\rm constant}$ \citep[e.g.,][]{2008ApJ...689.1234M}, we obtain
\begin{eqnarray}
{B_y}^2 \propto Pr_{m}^{1/2} R_{m}^{1/2}.\label{eq:21}
\end{eqnarray}
Since the Maxwell stress is proportional to the magnetic energy (Figure \ref{fig:bsqr-bstrs-vr}), equation (\ref{eq:21}) supports the simulation result of $w_{M} \propto Pr_{m}^{0.5-1.0}$.

{The linear analysis of equation (\ref{eq:21}) indicates that the saturation level depends on the magnetic Reynolds number as well as the magnetic Prandtl number.
However, the dependence on the magnetic Reynolds number is less pronounced than that on the magnetic Prandtl number within the explored parameter range in our visco-resistive simulation.
The magnetic Reynolds number dependence of the saturation level has been investigated by e.g., \cite{2000ApJ...530..464F} and \cite{2002ApJ...570..314S} by means of resistive MHD simulations.
\cite{2002ApJ...570..314S} have shown that the saturation level of the Maxwell stress decreases when $v_{A}^2/\eta \Omega \lsim 1$, and further, the level is nearly constant when $v_{A}^2/\eta \Omega \gsim 1$.
The magnetic Reynolds number dependence does not appear to obey a single power-law relation.
}

{ The rigorous linear analysis of the secondary parasitic instability under the primary MRI mode has shown that the fastest parasitic instability is associated with the KH mode, not the tearing mode, especially at high magnetic Prandtl number \citep{2009ApJ...698L..72P}.
By equating the growth rates of the primary MRI and the secondary parasitic instability, the saturation level is almost independent of the magnetic Prandtl number when the spatial domain is unlimited to permit the fastest primary and secondary instabilities.
However, they have shown that the saturation level weakly depends on the magnetic Prandtl number under the domain size of $(L_x,L_y,L_z) = 1\times 4 \times 1$.
The growth of the KH instability is suppressed by limiting the domain in the horizontal direction whereas the domain is sufficient in the azimuthal direction so as to permit the tearing instability. 
This situation may overestimate the relative contribution of the tearing mode to the parasitic instability, thereby causing the magnetic Prandtl number dependence.
In this regard, our interpretation of the simulation based on the simplified linear analysis follows the rigorous analysis by \cite{2009ApJ...698L..72P}.
As they have pointed out, it is necessary to use a sufficiently wide simulation domain (at least, $L_x,L_y > 2L_z$) to accommodate the fastest parasitic modes.
The (numerical) magnetic Prandtl number dependence of the saturation level may be moderated provided we use a wide domain in the horizontal direction so as to permit the fastest parasitic (KH) instability.
\cite{2008A&A...487....1B} have investigated the aspect ratio dependence of the saturation level of the MRI in the unstratified shearing box.
The channel flow that is well developed in the case with an aspect ratio of unity $L_x/L_z = 1$ disappears at an aspect ratio larger than unity, with which the KH instability is expected to disrupt the flow.
Insufficient domain size in the horizontal direction may lead to overestimating the saturation level of the MRI-induced turbulence.
}

{ The correlation between the magnetic Prandtl number and Maxwell stress as obtained by \cite{2007MNRAS.378.1471L} $(w_{M} \propto Pr_{m}^{0.25-0.5})$ is weaker than our result $(w_{M} \propto Pr_{m}^{0.5-1.0})$, indicating that other parameters affect the dependence.
This discrepancy may be due to the different range of the Reynolds number.
\cite{2007MNRAS.378.1471L} explored the magnetic Prandtl number dependence in a wide range of the Reynolds number $R_e = 200 - 6400$ by virtue of a pseudo-spectral incompressible code whereas our simulations are $R_e < 2000$.
The magnetic Prandtl number dependence in \cite{2007MNRAS.378.1471L} is more prominent for lower Reynolds number (Fig. 10 in their paper).
A similar tendency is found in the linear analysis by \cite{2009ApJ...698L..72P}, although the dependence is less pronounced than in the case of nonlinear simulation results.
}

\subsection{Gas Pressure Dependence}\label{sec:gas-press-depend}
{ The pressure-independent saturation level of the MRI is indicative of the incompressible turbulence in the intermediate to high pressure range.}
{On the other hand, the saturation level shows positive correlation with the gas pressure} at very low pressures (the leftmost symbols in Figures \ref{fig:strs-hllr-hll}(a), \ref{fig:strs-hll-hll-hllr-hllr}, and \ref{fig:strs_highres_ideal}) regardless of the choice of the scheme.
{ This may indicate the role of physical dissipation for the saturation of the MRI, which is irrelevant to numerical dissipation scales.}
The magnetic pressure is amplified by a factor of $\sim 100$ in our ideal MHD simulations regardless of the initial gas pressure.
Therefore, the plasma beta at the saturation is close to or smaller than unity in the lowest pressure case ($\beta = 10^2$).
The flow speed of $\sim v_{A}$ driven by the MRI becomes supersonic with $\beta \lsim 1$.
{ Consequently, compressibility may play an important role in the saturation of the MRI.
For example, slow-mode shocks contribute to the dissipation of the current and eventually lead to the decrease of the Maxwell stress with decreasing the pressure (increasing compressibility).}
{ As can be found in Figure \ref{fig:strs-n4-wrzh}(a), the time variation of the stress is significantly suppressed for the lowest pressure case.
This implies that a large-scale channel flow is inhibited from growing in large amplitudes by compressibility.
Subsequently, it will inhibit the secondary parasitic instability because its growth rate is proportional to the amplitude of the primary mode.
Alternatively, \cite{2009MNRAS.394..715L} have argued that non-linear multi-mode interactions by turbulent mixing, not the linear parasitic instability, is responsible for the saturation of the MRI in the case when the primary mode cannot reach large enough amplitudes.}
Although the compressible case ($\beta = 10^2$) in our simulation is not directly applicable to actual accretion disks because the vertical size is considerably larger than the pressure scale height, $L \gg \sqrt{2P/\rho}/\Omega$,
the simulation of the MRI with relatively low plasma beta is of interest to the fundamental study of compressible MHD turbulence.
The saturation mechanism of the compressible MRI-induced turbulence will be investigated in detail in future.

{ Although the saturation level of the MRI is found to be almost pressure-independent when the magnetic Prandtl number is constant, it does not necessarily hold true in actual plasma environments.
The saturation level may depend on the pressure because the magnetic Prandtl number can change with density, temperature, and hence pressure.
For example, the magnetic Prandtl number in weakly magnetized, collisional plasmas is estimated to be $Pr_m \sim 10^{-5} T^4/n$ for fully ionized cases and $Pr_m \sim 10^{7} T^2/n$ for partially ionized cases \citep{1962pfig.book.....S}.
\cite{2014MNRAS.441..681P} have considered the magnetic Prandtl number for radiation-dominated disks, $Pr_m \sim 10^{10} T^{11/2}/(\kappa n^{2})$, where $\kappa=(0.4+8 n T^{-7/2}) \; {\rm cm^2 g^{-1}}$ is the opacity for bound-free absorption in a partially ionized gas.
}

\section{SUMMARY}\label{sec:conclusion}
Magnetohydrodynamic (MHD) turbulence has been extensively studied via ideal MHD simulations.
In general, {ideal MHD simulations of turbulence with various schemes are not identical} because viscous and resistive dissipation scales differ among schemes.
We have investigated the saturation level of the MHD turbulence induced by the magnetorotational instability (MRI) by means of the ideal MHD simulation.
{ We use Godunov-type schemes with various reconstruction functions and Riemann solvers in order to assess effects of numerical dissipation}.
Upon revisiting the gas pressure dependence of the saturation level of the Maxwell stress as studies by \cite{2004ApJ...605..321S}, 
{ we have shown that the saturation level positively correlates with the pressure only at very low pressure, in which the MRI-driven flow becomes compressible.
Except for this case, we have failed to evaluate the gas pressure dependence of the saturation level because numerical viscous and resistive scales vary according to the pressure as well as the numerical scheme.}


We have estimated the numerical magnetic Prandtl number $Pr_{m}$ (=kinematic viscosity $\nu$/resistivity $\eta$), and we have shown that the saturation level is more sensitive to the numerical magnetic Prandtl number than the pressure.
{ Since the numerical magnetic Prandtl number itself depends on the pressure, the pressure dependence of the saturation level is indistinguishable from the numerical magnetic Prandtl number dependence.
Therefore, we conclude that the gas pressure dependence obtained by \cite{2004ApJ...605..321S} could be a consequence of the pressure-dependent numerical magnetic Prandtl number\footnote{In fact, the numerical magnetic Prandtl number for the scheme used in their paper is found to be an increasing function of the pressure. }.
We have also conducted the same runs with the ZEUS code, and we have confirmed that both the saturation level and the numerical magnetic Prandtl number are almost pressure-independent (Appendix ~\ref{sec:suppl-simul-using}).
One should recognize the numerical magnetic Prandtl number of the applied numerical code and its impact on the result.}
Also, the result strongly suggests the need for the explicit use of physical viscosity and resistivity to control the magnetic Prandtl number for the MRI-induced turbulence simulation as pointed out by e.g., \cite{2007A&A...476.1113F} and \cite{2008MNRAS.385.1036S}.
{ \cite{2007A&A...476.1113F} showed that solutions of the ideal MHD simulation of the MRI-induced turbulence with zero net magnetic flux depend on numerical resolution.
\cite{2008MNRAS.385.1036S} showed the numerical resolution dependence of the saturation level even with finite net flux cases, in which the minimum scale is limited by the critical wavelength of the MRI with respect to the finite ambient magnetic field, $\lambda = 2 \pi v_{A}/\sqrt{2 q} \Omega \propto B_0$.
In addition to them, we have revealed that the saturation level depends on the numerical magnetic Prandtl number, which is more sensitive to characteristics of the applied scheme rather than numerical resolution.}

Consequently, we have carried out a visco-resistive MHD simulation to investigate the parameter dependence of the saturation level of the MRI-induced turbulence.
The saturation level of the Maxwell stress depends on the physical magnetic Prandtl number.
{ It is almost independent of the gas pressure when the magnetic Prandtl number is constant and compressibility is weak.}
The positive correlation between the Maxwell stress and the magnetic Prandtl number may be an indication of the saturation mechanism.
{ We consider from linear analysis that the secondary tearing instability saturates the growth of the MRI, and consequently, it causes the magnetic Prandtl number dependence.}

\begin{acknowledgements}
We thank the anonymous referee for carefully reviewing our manuscript and giving insightful comments that improve the manuscript.
We also thank T. Miyoshi, Y. Masada, Y. Kawamura, and H. Sakaguchi for fruitful discussions and comments.
This work was partly supported by joint research project of ILE, Osaka University.
\end{acknowledgements}

\appendix
\section{IMPLEMENTATION OF THE MULTI-STATE RIEMANN SOLVER TO THE UPWIND CONSTRAINED TRANSPORT METHOD}\label{sec:appendix1}
We consider the two-dimensional ideal MHD equations,
\begin{eqnarray}
&& \pdif{\vect{U}}{t}+\pdif{\vect{F}}{x}+\pdif{\vect{G}}{y} = 0,\label{eq:28}\\
&& \pdif{B_x}{t}+\pdif{E_z}{y}=0,\label{eq:29}\\
&& \pdif{B_y}{t}-\pdif{E_z}{x}=0,\label{eq:30}
\end{eqnarray}
where $\vect{U} = (\rho,u_x,u_y,u_z,e,B_z)$ denotes conservative variables ($e$ the total energy), $\vect{F}(\vect{U},B_x,B_y)$ and $\vect{G}(\vect{U},B_x,B_y)$ the corresponding fluxes along $x$ and $y$ directions, and $E_z = u_yB_x-u_xB_y$ the electric field, respectively.
The extension to three dimension is straightforward.
The equations are discretized into a finite volume formulation as
\begin{eqnarray}
&& \dif{\vect{U}_{i,j}}{t} = -\frac{\vect{F}_{i+1/2,j}-\vect{F}_{i-1/2,j}}{\Delta x}-\frac{\vect{G}_{i,j+1/2}-\vect{G}_{i,j-1/2}}{\Delta y},\label{eq:31}\\
&& \dif{B_{x;i-1/2,j}}{t}=-\frac{E_{z;i-1/2,j+1/2}-E_{z;i-1/2,j-1/2}}{\Delta y},\label{eq:32}\\
&& \dif{B_{y;i,j-1/2}}{t}= \frac{E_{z;i+1/2,j-1/2}-E_{z;i-1/2,j-1/2}}{\Delta x},\label{eq:33}
\end{eqnarray}
where $\vect{U}_{i,j}$ defined at cell centers are averaged over the area and $B_{x;i-1/2,j},B_{y;i,j-1/2}$ defined at cell faces are averaged along the orthogonal line.
This is the staggered grid system employed by \cite{1988ApJ...332..659E}.

We solve the above equations with an upwind scheme based on Riemann solvers. 
First, we interpolate the in-plane magnetic fields to cell centers as cell-averaged representations. 
For example, the second, fourth, and sixth order approximations are respectively expressed as
\begin{eqnarray}
B_{y;i,j} &=& \frac{1}{2} \left( B_{y;i,j-1/2}+B_{y;i,j+1/2} \right),\label{eq:34}\\
B_{y;i,j} &=& \frac{1}{24}\left( -B_{y;i,j-3/2}+13B_{y;i,j-1/2}+13B_{y;i,j+1/2}-B_{y;i,j+3/2} \right),\label{eq:35}\\
B_{y;i,j} &=& \frac{1}{1440} \left( 11B_{y;i,j-5/2}-93B_{y;i,j-3/2}+802B_{y;i,j-1/2}\right.\nonumber \\
&& \left.+802B_{y;i,j+1/2}-93B_{y;i,j+3/2}+11B_{y;i,j+5/2} \right),\label{eq:36}
\end{eqnarray}
where we assume uniform grid spacing.
We adopt the sixth order approximation.
Since all the variables $\vect{W}_{i,j}=(\vect{U}_{i,j},B_{x;i,j},B_{y;i,j})$ are prepared at cell centers, they are interpolated along the $x$-direction by an arbitrary reconstruction function ${\cal F}_1$ with the degree of $2s+1$ to obtain the left and right states at right and left faces,
\begin{eqnarray}
\vect{W}_{i+1/2,j}^{Lx},\vect{W}_{i-1/2,j}^{Rx} \leftarrow {\cal F}_1\left( \vect{W}_{i-s,j} ,\dots, \vect{W}_{i,j} ,\dots, \vect{W}_{i+s,j}\right),\label{eq:37}
\end{eqnarray}
where the superscripts $Lx,Rx$ indicate the left and right states along the $x$-direction.
Note that the interpolation of $B_{x}$ (normal component) to cell faces is not required because it is already defined there.
Subsequently, we solve the Riemann problem at cell faces with an arbitrary one-dimensional Riemann solver ${\cal R}_1$ to determine the variable $\vect{W}_{i-1/2,j}$ as well as the upwind numerical flux $\vect{F}_{i-1/2,j}$.
For example, the HLLD approximate Riemann solver determines them to satisfy $\vect{F}_{\rm HLLD} = \vect{F}(\vect{W}_{\rm HLLD})$ \citep{2005JCoPh.208..315M}.

The above procedure is the one-dimensional reconstruction along the $x$-direction at $y_j$.
To determine the electric field $E_{z;i-1/2,j-1/2}$ at cell edges (numerical flux for the in-plane magnetic fields), we perform the same procedure along the $y$-direction at $x_{i-1/2}$.
The variable $\vect{W}_{i-1/2,j}$ is interpolated along the $y$-direction by a reconstruction function ${\cal F}_2$ to obtain their left and right states at right and left edges,
\begin{eqnarray}
\vect{W}_{i-1/2,j+1/2}^{Ly},\vect{W}_{i-1/2,j-1/2}^{Ry} \leftarrow {\cal F}_2\left( \vect{W}_{i-1/2,j-s} ,\dots, \vect{W}_{i-1/2,j} ,\dots, \vect{W}_{i-1/2,j+s}\right).\label{eq:40}
\end{eqnarray}
Subsequently, we solve the Riemann problem at cell edges with a Riemann solver ${\cal R}_2$ to determine the upwind electric field,
\begin{eqnarray}
E_{z;i-1/2,j-1/2} \leftarrow {\cal R}_2\left(\vect{W}_{i-1/2,j-1/2}^{Ly},\vect{W}_{i-1/2,j-1/2}^{Ry}\right).\label{eq:42}
\end{eqnarray}
Here, a multi-state Riemann solver is available for ${\cal R}_2$.

For symmetry, we perform the same procedure that involves the reconstruction along the $y$-direction at $x_i$ to obtain $\vect{W}_{i,j-1/2}$ followed by the reconstruction along the $x$-direction at $y_{j-1/2}$ to determine
\begin{eqnarray}
&& \vect{W}_{i+1/2,j-1/2}^{Lx},\vect{W}_{i-1/2,j-1/2}^{Rx} \leftarrow {\cal F}_2\left( \vect{W}_{i-s,j-1/2} ,\dots, \vect{W}_{i,j-1/2} ,\dots, \vect{W}_{i+s,j-1/2}\right),\label{eq:38}\\
&& E_{z;i-1/2,j-1/2} \leftarrow {\cal R}_2\left(\vect{W}_{i-1/2,j-1/2}^{Lx},\vect{W}_{i-1/2,j-1/2}^{Rx}\right).\label{eq:41}
\end{eqnarray}
Finally, we use the arithmetic average of Equations (\ref{eq:42}) and (\ref{eq:41}) as the numerical flux in Equations (\ref{eq:32}) and (\ref{eq:33}).
A similar method has been recently proposed to advance the Maxwell equation in kinetic plasma simulations \citep{2015CoPhC.187..137M}.

Our method is not identical to the upwind constrained transport method originally proposed by \cite{2004JCoPh.195...17L} in that they use a two-dimensional HLL Riemann solver whereas we use an arbitrary one-dimensional Riemann solver dimension-by-dimension. 
We consider that the essence is common between the two methods; calculation of the electric field at cell edges to retain the multi-dimensional upwind property.
Our method can easily incorporate multi-state Riemann solvers by virtue of successive one-dimensional reconstructions.

The accuracy of the magnetic field is determined by ${\cal F}_2,{\cal R}_2$ whereas that of the fluid is by ${\cal F}_1,{\cal R}_1$.
In this paper, we use various combinations of ${\cal F}_1,{\cal R}_1$ and ${\cal F}_2,{\cal R}_2$ to assess effects of numerical dissipation in the ideal MHD simulation of the MRI-induced turbulence.

\section{TRANSITION BETWEEN LAMINAR AND TURBULENT STATES}\label{sec:trans-betw-lamin}
As can be found in Table \ref{tab:sim_table_04}, the Maxwell stress is unexpectedly high at the cases with high viscosity.
For example, the Maxwell stress is nearly constant for the range of $\nu = \eta = (0.5-2)\times10^{-3}$ (V?-0-1,V?-1-1,V?-2-1), but it is suddenly increased at $\nu = \eta = 4\times10^{-3}$ (V?-4-1) and is fairly sensitive to the gas pressure.
Figure \ref{fig:case_prm=1} shows the time profile of the Maxwell stress and the vertical magnetic field energy $B_z^2/8 \pi$ normalized by the pressure in the run denoted as V4-4-1.
There are many quasi-periodic bursts with exponential growth as reported by \cite{2007MNRAS.378.1471L}.
Obviously, time averaging is meaningless in such a situation.
During the burst growth, the vertical magnetic field energy is nearly equal to the initial level and the growth rates of bursts are almost equal.
Therefore, the bursts are expected to be essentially identical to the initially unstable mode.
The periodic appearance of the burst suggests that the solution is laminar rather than turbulent.


The origin of the burst is understood from linear theory.
The dispersion relation \cite[e.g.,][]{2007MNRAS.378.1471L} is analytically solved at $Pr_m = 1$ as
\begin{eqnarray}
\frac{\gamma}{\Omega} = -\frac{1}{S} \left(\frac{k v_{A}}{\Omega}\right)^2 + \left[ \sqrt{4 \left(\frac{k v_{A}}{\Omega}\right)^2 + \left(2-q\right)^2} - \left(\frac{k v_{A}}{\Omega}\right)^2 - \left(2-q\right)\right]^{1/2},\label{eq:22}
\end{eqnarray}
where $S = v_{A}^2/\eta \Omega (=v_{A}^2/\nu \Omega)$ denotes the Lundquist number.
The growth rate and the wavenumber of the fastest growing mode (FGM) are approximated as 
\begin{eqnarray}
\left(\frac{\gamma}{\Omega}\right)_{\rm FGM} \simeq \frac{q}{2} \left(1-\frac{2-q}{2S}\right) \;\;\; {\rm at} \;\;\; \left(\frac{k v_{A}}{\Omega}\right)_{\rm FGM}^2 \simeq \frac{1}{1+2q/S} - \frac{\left(2-q\right)^2}{4},\label{eq:23}
\end{eqnarray}
for $S \gg 1$, and 
\begin{eqnarray}
\left(\frac{\gamma}{\Omega}\right)_{\rm FGM} \simeq \frac{S}{2}\left(\sqrt{\frac{2q}{2-q}}-1\right) \;\;\; {\rm at} \;\;\; \left(\frac{k v_{A}}{\Omega}\right)_{\rm FGM} \simeq \frac{S}{\sqrt{2}},\label{eq:24}
\end{eqnarray}
for $S \ll 1$.
The critical wavenumber for the instability is $k_{\rm crit}v_{A}/\Omega = \sqrt{2q}$ for $S \gg 1$ (ideal limit) and $k_{\rm crit}v_{A}/\Omega = \sqrt{q/(2-q)}S$ for $S \ll 1$.

The instability condition is characterized by a combination of the Lundquist number $S$ and the minimum wavenumber limited by the vertical size $k_{\rm min} = 2 \pi /L$.
When the minimum wavenumber is considerably smaller than that of the FGM, 
the system involves multiple unstable modes, and subsequently, the turbulence is successfully sustained by the MRI. 
On the other hand, only the longest wavelength mode grows when the minimum wavenumber is larger than that of the FGM and smaller than the critical wavenumber. 
In such a situation, we observe only the largest laminar flow that is marginally unstable \citep{2007MNRAS.378.1471L}.
In the runs at $Pr_m=1$ (V?-0-1,V?-1-1,V?-2-1,V?-4-1), the condition for the laminar regime for $S \ll 1$,
\begin{eqnarray}
\frac{1}{\sqrt{2}} < \frac{k_{\rm min} \nu}{v_{A}} < \sqrt{\frac{q}{2-q}}=\sqrt{3},\label{eq:25} 
\end{eqnarray}
is satisfied only with $\nu=4\times 10^{-3}$ (V?-4-1), in which the quasi-periodic burst with $k=k_{\rm min}$ is actually observed.
The growth rate of $\gamma = 0.1\Omega$ is in good agreement with the exponential growth of bursts.

The laminar regime also appears in the high magnetic Prandtl number cases (e.g., runs V?-4-2 and V?-4-4).
This is expected as per linear theory in the viscous limit $(\nu \gg v_A^2/\Omega,\eta=0)$.
The wavenumber of the fastest growing mode decreases with the viscosity, $k_{\rm FGM} = \sqrt{2(2-q)\Omega/\nu}$, but the critical wavenumber remains unchanged, $k_{\rm crit} = \sqrt{2q}\Omega/v_{A}$ \citep{2008ApJ...684..498P,2008ApJ...689.1234M}.
Therefore, the largest flow is unstable regardless of the viscosity.
The condition for the laminar regime in the viscous limit is given by
\begin{eqnarray}
 \frac{v_{A}}{\sqrt{2q} \Omega} < \frac{L}{2 \pi} < \sqrt{\frac{\nu}{2(2-q)\Omega}}.\label{eq:27}
\end{eqnarray}
The exponential burst is most likely to be observed at high viscosity.
This is not the case in the resistive limit ($\nu = 0, \eta \gg v_A^2/\Omega$) because the critical wavenumber as well as the wavenumber of the fastest growing mode decrease with the resistivity \citep{1999ApJ...515..776S}.

\section{LINEAR ANALYSIS OF VISCO-RESISTIVE KELVIN-HELMHOLTZ AND TEARING INSTABILITIES}\label{sec:line-analys-visco-KH-tear}
In the incompressible visco-resistive MHD, the linearized equations of the KH instability \citep{1961hhs..book.....C} and the tearing instability \citep{1963PhFl....6..459F} are respectively written as
\begin{eqnarray}
\left\{
\begin{array}{l}
\gamma_{\rm KH} A = ik \left[ \cos \theta \psi - FA\right] + R_{m}^{-1} \Delta A,\\
\gamma_{\rm KH} \Delta \psi = ik \left[F'' \psi - F \Delta \psi + \left(\frac{v_{A}}{v_{0}}\right)^2 \cos \theta \Delta A\right] + R_{e}^{-1} \Delta \Delta \psi,
\end{array} 
\right.\label{eq:18}
\end{eqnarray}
and
\begin{eqnarray}
\left\{
\begin{array}{l}
\gamma_{\rm tear} A = F \psi + R_{m}^{-1} \Delta A,\\
\gamma_{\rm tear} \Delta \psi = k^2\left(F''A-F\Delta A\right) + R_{e}^{-1} \Delta \Delta \psi,
\end{array} 
\right.\label{eq:19}
\end{eqnarray}
where $A$ and $\psi$ denote the flux and stream functions, $R_e$ and $R_m$ the Reynolds and magnetic Reynolds numbers, $\Delta = -k^2+d^2/dz^2$, $F=\tanh(z)$ the profile of the background shear flow (KH) or magnetic field (tearing), and $F''=d^2F/dz^2$.
The parameters $v_{A}$ and $v_{0}$ denote the {\Alfven} velocity and the half velocity jump, $\theta$ denotes the angle between the shear flow and magnetic field, and we assume that the KH wave is parallel to the shear flow.
{The in-plane magnetic field is included in the KH instability, and the background flow is ignored in the tearing instability for simplicity.}
In equation (\ref{eq:18}), the variables are normalized by $v_0$ and the width of the shear layer $\delta$.
{The Reynolds and magnetic Reynolds numbers are expressed as $R_e=v_0 \delta/\nu$ and $R_m=v_0 \delta/\eta$, respectively.}
In equation (\ref{eq:19}), the variables are normalized by the {\Alfven} velocity and the width of the current layer $\delta$, thus the Reynolds and magnetic Reynolds numbers are expressed as $R_e=v_A \delta/\nu$ and $R_m=v_A \delta/\eta$, respectively.
{We numerically obtain the eigenvalues and eigenfunctions with the boundary condition $\Delta A = \Delta \psi = 0$ at $|z|=\pm 10$.}

\section{SUPPLEMENTARY SIMULATIONS USING THE ZEUS CODE}\label{sec:suppl-simul-using}
Using the ZEUS code \citep{1992ApJS...80..753S,1992ApJS...80..791S}, we ran a set of ideal MHD simulations to investigate the gas pressure dependence of the saturation level of MRI.
The parameters are the same as those in the text except for the net vertical field $B_0$, which is larger by a factor of $\sqrt{4\pi}$. (There is no
particular reason for this choice of $B_0$.)
Figure \ref{fig:zeus}(a) shows the Maxwell stress (averaged from 20 to 100 orbits) vs. gas pressure.
The Maxwell stress is almost independent of gas pressure except the smallest pressure case. 
Then we evaluated the numerical magnetic Prandtl numbers for these simulations in the same manner as explained in the text
except that each number is calculated from a statistical average of four simulations restarting from different epochs (20, 40, 60, and 80 orbits).
Figure \ref{fig:zeus}(b) shows that the numerical magnetic Prandtl number is roughly constant regardless of gas pressure.
These results support the statement in the text that the saturation level does not depend on gas pressure (except for very low plasma beta cases)
when (numerical) magnetic Prandtl number is constant.

\bibliographystyle{apj}                                                       

\begin{thebibliography}{54}
\expandafter\ifx\csname natexlab\endcsname\relax\def\natexlab#1{#1}\fi

\bibitem[{{Balbus} \& {Hawley}(1991)}]{1991ApJ...376..214B}
{Balbus}, S.~A., \& {Hawley}, J.~F. 1991, \apj, 376, 214

\bibitem[{{Balbus} \& {Hawley}(1998)}]{1998RvMP...70....1B}
---. 1998, Reviews of Modern Physics, 70, 1

\bibitem[{{Balsara} \& {Shu}(2000)}]{2000JCoPh.160..405B}
{Balsara}, D.~S., \& {Shu}, C.-W. 2000, Journal of Computational Physics, 160,
  405

\bibitem[{{Bodo} {et~al.}(2011){Bodo}, {Cattaneo}, {Ferrari}, {Mignone}, \&
  {Rossi}}]{2011ApJ...739...82B}
{Bodo}, G., {Cattaneo}, F., {Ferrari}, A., {Mignone}, A., \& {Rossi}, P. 2011,
  \apj, 739, 82

\bibitem[{{Bodo} {et~al.}(2008){Bodo}, {Mignone}, {Cattaneo}, {Rossi}, \&
  {Ferrari}}]{2008A&A...487....1B}
{Bodo}, G., {Mignone}, A., {Cattaneo}, F., {Rossi}, P., \& {Ferrari}, A. 2008,
  \aap, 487, 1

\bibitem[{{Borges} {et~al.}(2008){Borges}, {Carmona}, {Costa}, \&
  {Don}}]{2008JCoPh.227.3191B}
{Borges}, R., {Carmona}, M., {Costa}, B., \& {Don}, W.~S. 2008, Journal of
  Computational Physics, 227, 3191

\bibitem[{{Chandrasekhar}(1961)}]{1961hhs..book.....C}
{Chandrasekhar}, S. 1961, {Hydrodynamic and hydromagnetic stability} (Oxford:
  Clarendon)

\bibitem[{{Del Zanna} {et~al.}(2007){Del Zanna}, {Zanotti}, {Bucciantini}, \&
  {Londrillo}}]{2007A&A...473...11D}
{Del Zanna}, L., {Zanotti}, O., {Bucciantini}, N., \& {Londrillo}, P. 2007,
  \aap, 473, 11

\bibitem[{{Evans} \& {Hawley}(1988)}]{1988ApJ...332..659E}
{Evans}, C.~R., \& {Hawley}, J.~F. 1988, \apj, 332, 659

\bibitem[{{Fleming} {et~al.}(2000){Fleming}, {Stone}, \&
  {Hawley}}]{2000ApJ...530..464F}
{Fleming}, T.~P., {Stone}, J.~M., \& {Hawley}, J.~F. 2000, \apj, 530, 464

\bibitem[{{Fromang} {et~al.}(2006){Fromang}, {Hennebelle}, \&
  {Teyssier}}]{2006A&A...457..371F}
{Fromang}, S., {Hennebelle}, P., \& {Teyssier}, R. 2006, \aap, 457, 371

\bibitem[{{Fromang} \& {Papaloizou}(2007)}]{2007A&A...476.1113F}
{Fromang}, S., \& {Papaloizou}, J. 2007, \aap, 476, 1113

\bibitem[{{Fromang} {et~al.}(2007){Fromang}, {Papaloizou}, {Lesur}, \&
  {Heinemann}}]{2007A&A...476.1123F}
{Fromang}, S., {Papaloizou}, J., {Lesur}, G., \& {Heinemann}, T. 2007, \aap,
  476, 1123

\bibitem[{{Furth} {et~al.}(1963){Furth}, {Killeen}, \&
  {Rosenbluth}}]{1963PhFl....6..459F}
{Furth}, H.~P., {Killeen}, J., \& {Rosenbluth}, M.~N. 1963, Physics of Fluids,
  6, 459

\bibitem[{{Goodman} \& {Xu}(1994)}]{1994ApJ...432..213G}
{Goodman}, J., \& {Xu}, G. 1994, \apj, 432, 213

\bibitem[{{Gressel} \& {Ziegler}(2007)}]{2007CoPhC.176..652G}
{Gressel}, O., \& {Ziegler}, U. 2007, Computer Physics Communications, 176, 652

\bibitem[{{Guan} {et~al.}(2009){Guan}, {Gammie}, {Simon}, \&
  {Johnson}}]{2009ApJ...694.1010G}
{Guan}, X., {Gammie}, C.~F., {Simon}, J.~B., \& {Johnson}, B.~M. 2009, \apj,
  694, 1010

\bibitem[{{Harten} {et~al.}(1983){Harten}, {Lax}, \& {van
  Leer}}]{1983SIAMrev.25..35M}
{Harten}, A., {Lax}, P., \& {van Leer}, B. 1983, SIAM review, 35, 35

\bibitem[{{Hawley} {et~al.}(1995){Hawley}, {Gammie}, \&
  {Balbus}}]{1995ApJ...440..742H}
{Hawley}, J.~F., {Gammie}, C.~F., \& {Balbus}, S.~A. 1995, \apj, 440, 742

\bibitem[{{Hawley} {et~al.}(1996){Hawley}, {Gammie}, \&
  {Balbus}}]{1996ApJ...464..690H}
---. 1996, \apj, 464, 690

\bibitem[{{Jiang} \& {Shu}(1996)}]{1996JCoPh.126..202J}
{Jiang}, G., \& {Shu}, C. 1996, Journal of Computational Physics, 126, 202

\bibitem[{{Kritsuk} {et~al.}(2011){Kritsuk}, {Nordlund}, {Collins}, {Padoan},
  {Norman}, {Abel}, {Banerjee}, {Federrath}, {Flock}, {Lee}, {Li},
  {M{\"u}ller}, {Teyssier}, {Ustyugov}, {Vogel}, \& {Xu}}]{2011ApJ...737...13K}
{Kritsuk}, A.~G., {Nordlund}, {\AA}., {Collins}, D., {Padoan}, P., {Norman},
  M.~L., {Abel}, T., {Banerjee}, R., {Federrath}, C., {Flock}, M., {Lee}, D.,
  {Li}, P.~S., {M{\"u}ller}, W.-C., {Teyssier}, R., {Ustyugov}, S.~D., {Vogel},
  C., \& {Xu}, H. 2011, \apj, 737, 13

\bibitem[{{Latter} {et~al.}(2009){Latter}, {Lesaffre}, \&
  {Balbus}}]{2009MNRAS.394..715L}
{Latter}, H.~N., {Lesaffre}, P., \& {Balbus}, S.~A. 2009, \mnras, 394, 715

\bibitem[{{Lesur} \& {Longaretti}(2007)}]{2007MNRAS.378.1471L}
{Lesur}, G., \& {Longaretti}, P.-Y. 2007, \mnras, 378, 1471

\bibitem[{{Londrillo} \& {Del Zanna}(2000)}]{2000ApJ...530..508L}
{Londrillo}, P., \& {Del Zanna}, L. 2000, \apj, 530, 508

\bibitem[{{Londrillo} \& {del Zanna}(2004)}]{2004JCoPh.195...17L}
{Londrillo}, P., \& {del Zanna}, L. 2004, Journal of Computational Physics,
  195, 17

\bibitem[{{Masada} \& {Sano}(2008)}]{2008ApJ...689.1234M}
{Masada}, Y., \& {Sano}, T. 2008, \apj, 689, 1234

\bibitem[{{Mignone} {et~al.}(2007){Mignone}, {Bodo}, {Massaglia}, {Matsakos},
  {Tesileanu}, {Zanni}, \& {Ferrari}}]{2007ApJS..170..228M}
{Mignone}, A., {Bodo}, G., {Massaglia}, S., {Matsakos}, T., {Tesileanu}, O.,
  {Zanni}, C., \& {Ferrari}, A. 2007, \apjs, 170, 228

\bibitem[{{Minoshima} {et~al.}(2015){Minoshima}, {Matsumoto}, \&
  {Amano}}]{2015CoPhC.187..137M}
{Minoshima}, T., {Matsumoto}, Y., \& {Amano}, T. 2015, Computer Physics
  Communications, 187, 137

\bibitem[{{Miyoshi} \& {Kusano}(2005)}]{2005JCoPh.208..315M}
{Miyoshi}, T., \& {Kusano}, K. 2005, Journal of Computational Physics, 208, 315

\bibitem[{{Miyoshi} \& {Kusano}(2008)}]{2008ASPC..385..279M}
{Miyoshi}, T., \& {Kusano}, K. 2008, in Astronomical Society of the Pacific
  Conference Series, Vol. 385, Numerical Modeling of Space Plasma Flows, ed.
  N.~V. {Pogorelov}, E.~{Audit}, \& G.~P. {Zank}, 279

\bibitem[{{Pessah} \& {Chan}(2008)}]{2008ApJ...684..498P}
{Pessah}, M.~E., \& {Chan}, C.-k. 2008, \apj, 684, 498

\bibitem[{{Pessah} {et~al.}(2007){Pessah}, {Chan}, \&
  {Psaltis}}]{2007ApJ...668L..51P}
{Pessah}, M.~E., {Chan}, C.-k., \& {Psaltis}, D. 2007, \apjl, 668, L51

\bibitem[{{Pessah} \& {Goodman}(2009)}]{2009ApJ...698L..72P}
{Pessah}, M.~E., \& {Goodman}, J. 2009, \apjl, 698, L72

\bibitem[{{Porcelli}(1987)}]{1987PhFl...30.1734P}
{Porcelli}, F. 1987, Physics of Fluids, 30, 1734

\bibitem[{{Potter} \& {Balbus}(2014)}]{2014MNRAS.441..681P}
{Potter}, W.~J., \& {Balbus}, S.~A. 2014, \mnras, 441, 681

\bibitem[{{Roe}(1981)}]{1981JCoPh..43..357R}
{Roe}, P.~L. 1981, Journal of Computational Physics, 43, 357

\bibitem[{{Sano} \& {Inutsuka}(2001)}]{2001ApJ...561L.179S}
{Sano}, T., \& {Inutsuka}, S.-i. 2001, \apjl, 561, L179

\bibitem[{{Sano} {et~al.}(2004){Sano}, {Inutsuka}, {Turner}, \&
  {Stone}}]{2004ApJ...605..321S}
{Sano}, T., {Inutsuka}, S.-i., {Turner}, N.~J., \& {Stone}, J.~M. 2004, \apj,
  605, 321

\bibitem[{{Sano} \& {Miyama}(1999)}]{1999ApJ...515..776S}
{Sano}, T., \& {Miyama}, S.~M. 1999, \apj, 515, 776

\bibitem[{{Sano} \& {Stone}(2002)}]{2002ApJ...570..314S}
{Sano}, T., \& {Stone}, J.~M. 2002, \apj, 570, 314

\bibitem[{{Schekochihin} {et~al.}(2004){Schekochihin}, {Cowley}, {Taylor},
  {Maron}, \& {McWilliams}}]{2004ApJ...612..276S}
{Schekochihin}, A.~A., {Cowley}, S.~C., {Taylor}, S.~F., {Maron}, J.~L., \&
  {McWilliams}, J.~C. 2004, \apj, 612, 276

\bibitem[{{Shakura} \& {Sunyaev}(1973)}]{1973A&A....24..337S}
{Shakura}, N.~I., \& {Sunyaev}, R.~A. 1973, \aap, 24, 337

\bibitem[{{Shu} \& {Osher}(1988)}]{1988JCoPh..77..439S}
{Shu}, C., \& {Osher}, S. 1988, Journal of Computational Physics, 77, 439

\bibitem[{{Silvers}(2008)}]{2008MNRAS.385.1036S}
{Silvers}, L.~J. 2008, \mnras, 385, 1036

\bibitem[{{Spitzer}(1962)}]{1962pfig.book.....S}
{Spitzer}, L. 1962, {Physics of Fully Ionized Gases}

\bibitem[{{Stone} \& {Gardiner}(2010)}]{2010ApJS..189..142S}
{Stone}, J.~M., \& {Gardiner}, T.~A. 2010, \apjs, 189, 142

\bibitem[{{Stone} {et~al.}(2008){Stone}, {Gardiner}, {Teuben}, {Hawley}, \&
  {Simon}}]{2008ApJS..178..137S}
{Stone}, J.~M., {Gardiner}, T.~A., {Teuben}, P., {Hawley}, J.~F., \& {Simon},
  J.~B. 2008, \apjs, 178, 137

\bibitem[{{Stone} \& {Norman}(1992{\natexlab{a}})}]{1992ApJS...80..753S}
{Stone}, J.~M., \& {Norman}, M.~L. 1992{\natexlab{a}}, \apjs, 80, 753

\bibitem[{{Stone} \& {Norman}(1992{\natexlab{b}})}]{1992ApJS...80..791S}
---. 1992{\natexlab{b}}, \apjs, 80, 791

\bibitem[{{Suresh} \& {Huynh}(1997)}]{1997JCoPh.136...83S}
{Suresh}, A., \& {Huynh}, H.~T. 1997, Journal of Computational Physics, 136, 83

\bibitem[{{Titarev} \& {Toro}(2004)}]{2004JCoPh.201..238T}
{Titarev}, V.~A., \& {Toro}, E.~F. 2004, Journal of Computational Physics, 201,
  238

\bibitem[{{Ziegler}(2004)}]{2004JCoPh.196..393Z}
{Ziegler}, U. 2004, Journal of Computational Physics, 196, 393

\bibitem[{{Ziegler}(2008)}]{2008CoPhC.179..227Z}
---. 2008, Computer Physics Communications, 179, 227

\end{thebibliography}

\gdef\thefigure{\arabic{figure}}

\clearpage

\clearpage

\clearpage

\clearpage

\clearpage

\clearpage

\clearpage

\clearpage

\clearpage

\clearpage

\clearpage


\clearpage
\begin{figure}[p]
\epsscale{1.0}
\plotone{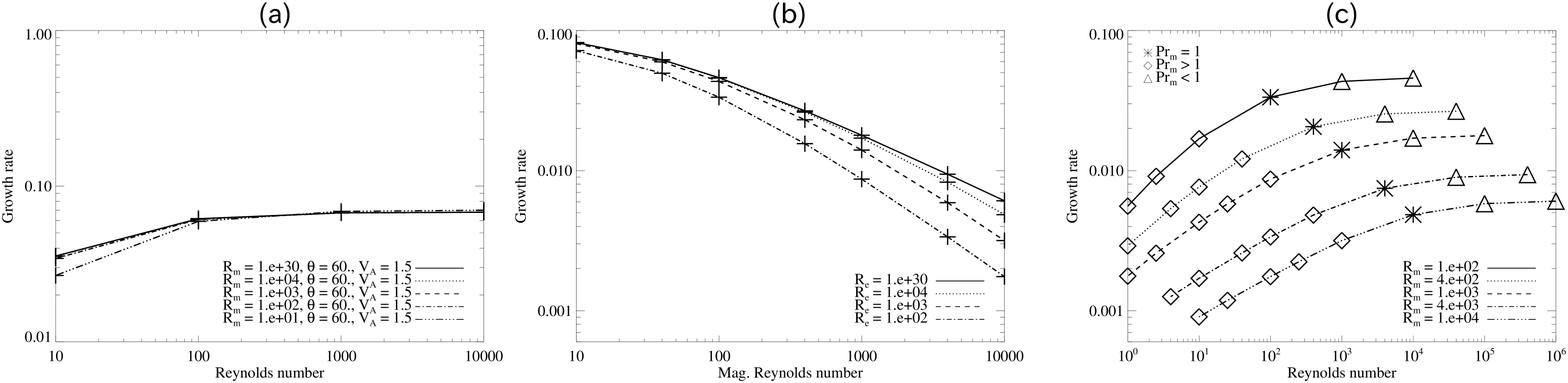}
\caption{Linear growth rate in the incompressible visco-resistive MHD. (a) The Kelvin-Helmholtz instability as a function of the Reynolds number. (b,c) The tearing instability as a function of the magnetic Reynolds number and the Reynolds number.}
\label{fig:kh_linear}
\end{figure}

\clearpage

\clearpage
\begin{figure}[p]
\epsscale{1.0}
\plotone{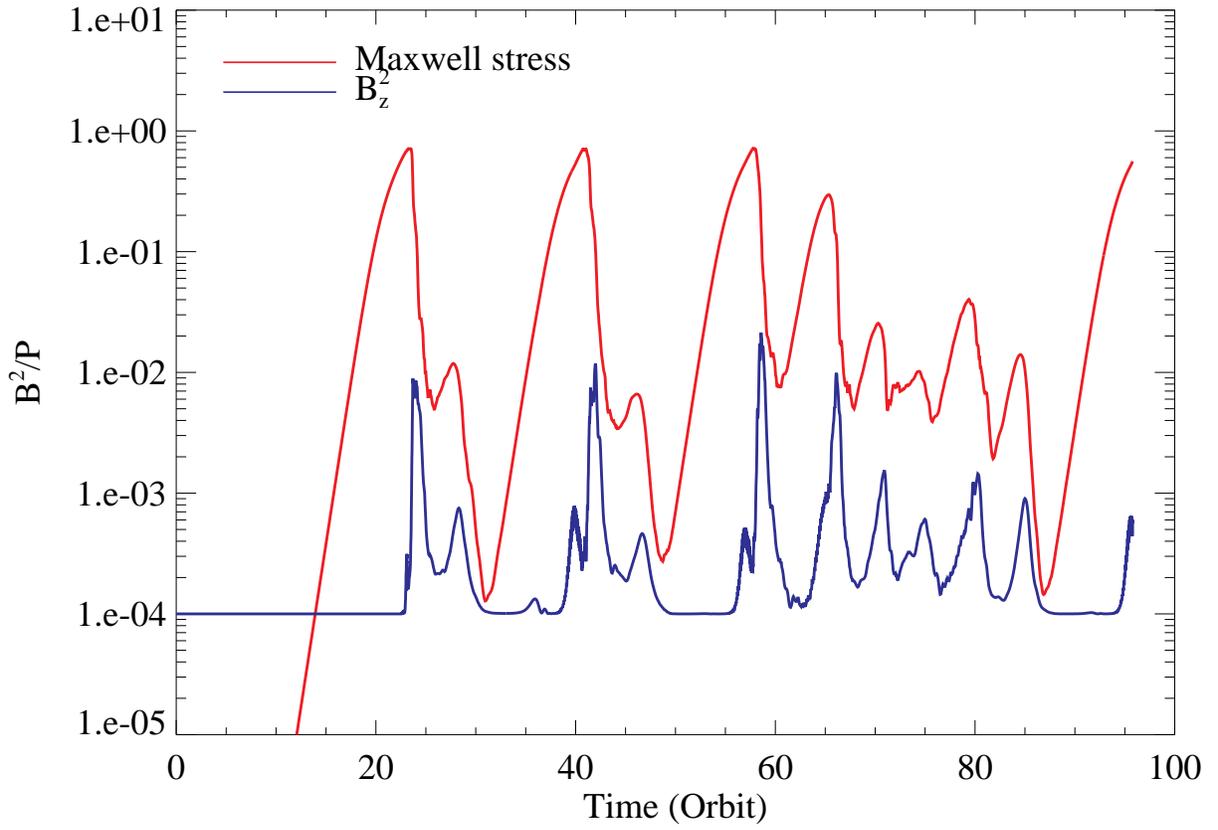}
\caption{Time profile of the Maxwell stress (red line) and the vertical magnetic field energy (blue line) in the run V4-4-1.}
\label{fig:case_prm=1}
\end{figure}

\clearpage
\begin{figure}[p]
\epsscale{1.0}
\plotone{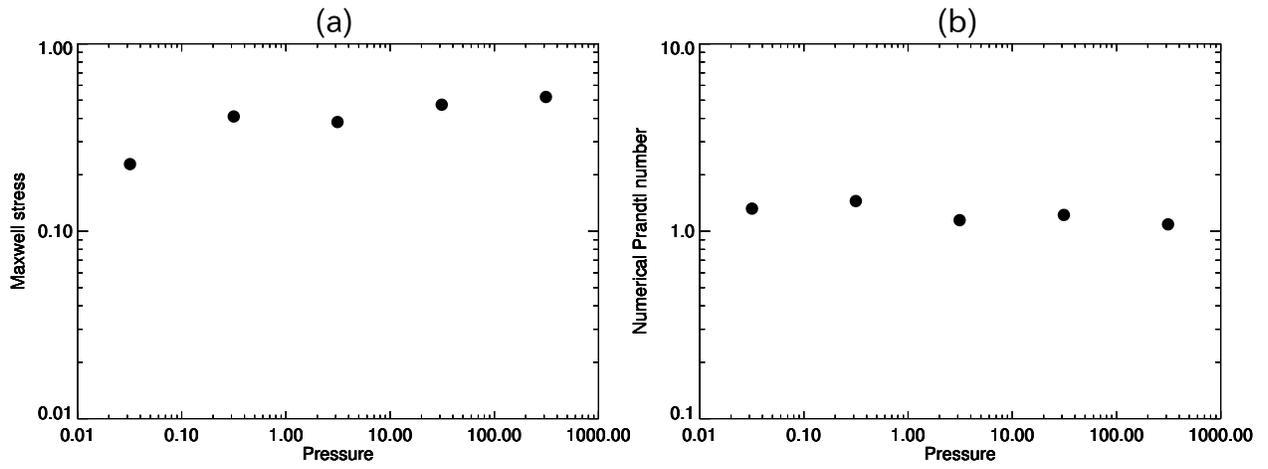}
\caption{Pressure dependence of the saturation level of MRI (a) and the numerical magnetic Prandtl number (b) based on ideal MHD simulations using the ZEUS code.}
\label{fig:zeus}
\end{figure}



\clearpage
\begin{center}
\begin{table}[htbp]
\caption{Ideal MHD simulations of the MRI. The HLLR and HLL Riemann solvers are used for the fluid and magnetic field updates, respectively.}
\label{tab:sim_table_01}
\scalebox{0.5}{
\begin{tabular}{ccccccccccc}
\toprule
(1) Model & (2) F1 & (3) F2 & (4) $\beta$ & (5) $P_0$ & (6) $E_{M} \times 10^{2}$ & (7) $E_{K} \times 10^{2}$ & (8) $w_{M} \times 10^{2}$ & (9) $w_{R} \times 10^{2}$ & (10) $\alpha \times 10^{2}$ & (11) $Pr_{m,{\rm num}}$ \\ \midrule
F1-HLLR-F2-HLL&&&&&&&&&&\\ \midrule
I2-WRWH & WENO & WENO & $10^2$ & $3.125\times 10^{-2}$ & $4.05$ & $1.23$ & $1.46$& $0.357 $ & $58.1$ & 0.76 \\
I3-WRWH & WENO & WENO & $10^3$ & $3.125\times 10^{-1}$ & $6.49$ & $2.60$ & $2.76$& $0.745 $ & $11.2$ & 0.43 \\
I4-WRWH & WENO & WENO & $10^4$ & $3.125\times 10^{0}$  & $5.23$ & $2.43$ & $2.32$& $0.642 $ & $0.948$ & 0.51 \\
I5-WRWH & WENO & WENO & $10^5$ & $3.125\times 10^{1}$  & $4.19$ & $1.72$ & $1.81$& $0.411 $ & $0.0710$ & 0.45 \\
I6-WRWH & WENO & WENO & $10^6$ & $3.125\times 10^{2}$  & $2.93$ & $1.03$ & $1.07$& $0.258 $ & $0.00426$ & 0.36 \\
I2-WRZH & WENO & WZMP & $10^2$ & $3.125\times 10^{-2}$ & $4.82$ & $1.42$ & $1.75$& $0.438 $ & $70.3$ & 1.6 \\
I3-WRZH & WENO & WZMP & $10^3$ & $3.125\times 10^{-1}$ & $8.07$ & $3.21$ & $3.52$& $0.946 $ & $14.4$ & 0.88 \\
I4-WRZH & WENO & WZMP & $10^4$ & $3.125\times 10^{0}$  & $7.63$ & $3.25$ & $3.47$& $0.871 $ & $1.39$ & 1.4 \\
I5-WRZH & WENO & WZMP & $10^5$ & $3.125\times 10^{1}$  & $9.28$ & $3.61$ & $4.31$& $0.957 $ & $0.169$ & 1.4 \\
I6-WRZH & WENO & WZMP & $10^6$ & $3.125\times 10^{2}$  & $12.5$ & $4.30$ & $5.53$& $1.21 $ & $0.0215$ & 0.84 \\
I3-ZRZH & WZMP & WZMP & $10^3$ & $3.125\times 10^{-1}$ & $7.13$ & $3.05$ & $3.07$& $0.851 $ & $12.7$ & 0.51 \\
I4-ZRZH & WZMP & WZMP & $10^4$ & $3.125\times 10^{0}$  & $5.28$ & $2.77$ & $2.45$& $0.684 $ & $1.00$ & 0.82 \\
I5-ZRZH & WZMP & WZMP & $10^5$ & $3.125\times 10^{1}$  & $5.38$ & $2.52$ & $2.47$& $0.580 $ & $0.0977$ & 0.74 \\
I6-ZRZH & WZMP & WZMP & $10^6$ & $3.125\times 10^{2}$  & $4.32$ & $2.04$ & $1.95$& $0.493 $ & $0.00782$ & 0.48 \\
\bottomrule
\end{tabular}
}
\end{table}
\end{center}

\clearpage
\begin{center}
\begin{table}[htbp]
\caption{Ideal MHD simulations of the MRI. The HLL Riemann solver is used for both the fluid and magnetic field updates.}
\label{tab:sim_table_02}
\scalebox{0.5}{
\begin{tabular}{ccccccccccc}
\toprule
(1) Model & (2) F1 & (3) F2 & (4) $\beta$ & (5) $P_0$ & (6) $E_{M} \times 10^{2}$ & (7) $E_{K} \times 10^{2}$ & (8) $w_{M} \times 10^{2}$ & (9) $w_{R} \times 10^{2}$ & (10) $\alpha \times 10^{2}$ & (11) $Pr_{m,{\rm num}}$ \\ \midrule
F1-HLL-F2-HLL&&&&&&&&&&\\ \midrule
I2-WHWH & WENO & WENO & $10^2$ & $3.125\times 10^{-2}$ & $2.01$ & $1.04$ & $0.848$& $0.307 $ & $37.0$ & 0.95 \\
I3-WHWH & WENO & WENO & $10^3$ & $3.125\times 10^{-1}$ & $6.55$ & $2.26$ & $2.79$& $0.655 $ & $11.0$ & 0.91 \\
I4-WHWH & WENO & WENO & $10^4$ & $3.125\times 10^{0}$  & $7.58$ & $2.84$ & $3.50$& $0.781 $ & $1.37$ & 0.86 \\
I5-WHWH & WENO & WENO & $10^5$ & $3.125\times 10^{1}$  & $6.93$ & $1.96$ & $2.97$& $0.496 $ & $0.111$ & 0.94 \\
I6-WHWH & WENO & WENO & $10^6$ & $3.125\times 10^{2}$  & $5.78$ & $1.31$ & $2.20$& $0.359 $ & $0.00819$ & 0.92 \\
I2-WHZH & WENO & WZMP & $10^2$ & $3.125\times 10^{-2}$ & $3.19$ & $1.15$ & $1.27$& $0.349 $ & $51.8$ & 1.3 \\
I3-WHZH & WENO & WZMP & $10^3$ & $3.125\times 10^{-1}$ & $8.24$ & $2.77$ & $3.58$& $0.846 $ & $14.2$ & 1.2 \\
I4-WHZH & WENO & WZMP & $10^4$ & $3.125\times 10^{0}$  & $9.02$ & $3.04$ & $4.16$& $0.877 $ & $1.61$ & 1.4 \\
I5-WHZH & WENO & WZMP & $10^5$ & $3.125\times 10^{1}$  & $10.8$ & $3.03$ & $5.06$& $0.832 $ & $0.189$ & 1.3 \\
I6-WHZH & WENO & WZMP & $10^6$ & $3.125\times 10^{2}$  & $11.7$ & $2.59$ & $5.07$& $0.782 $ & $0.0187$ & 1.6 \\
I3-ZHZH & WZMP & WZMP & $10^3$ & $3.125\times 10^{-1}$ & $8.43$ & $3.22$ & $3.66$& $0.922 $ & $14.7$ & 0.90 \\
I4-ZHZH & WZMP & WZMP & $10^4$ & $3.125\times 10^{0}$  & $10.0$ & $3.98$ & $4.61$& $1.01 $ & $1.80$ & 1.1 \\
I5-ZHZH & WZMP & WZMP & $10^5$ & $3.125\times 10^{1}$  & $6.30$ & $2.54$ & $3.02$& $0.579 $ & $0.115$ & 1.3 \\
I6-ZHZH & WZMP & WZMP & $10^6$ & $3.125\times 10^{2}$  & $6.60$ & $2.43$ & $3.10$& $0.582 $ & $0.0118$ & 0.99 \\
\bottomrule
\end{tabular}
}
\end{table}
\end{center}

\clearpage
\begin{center}
\begin{table}[htbp]
\caption{Ideal MHD simulations of the MRI. The HLLR Riemann solver is used for both the fluid and magnetic field updates.}
\label{tab:sim_table_03}
\scalebox{0.5}{
\begin{tabular}{ccccccccccc}
\toprule
(1) Model & (2) F1 & (3) F2 & (4) $\beta$ & (5) $P_0$ & (6) $E_{M} \times 10^{2}$ & (7) $E_{K} \times 10^{2}$ & (8) $w_{M} \times 10^{2}$ & (9) $w_{R} \times 10^{2}$ & (10) $\alpha \times 10^{2}$ & (11) $Pr_{m,{\rm num}}$ \\ \midrule
F1-HLLR-F2-HLLR&&&&&&&&&&\\ \midrule
I2-WRWR & WENO & WENO & $10^2$ & $3.125\times 10^{-2}$ & $6.32$ & $1.68$ & $2.27$& $0.535 $ & $90.4$ & 2.7 \\
I3-WRWR & WENO & WENO & $10^3$ & $3.125\times 10^{-1}$ & $11.4$ & $3.87$ & $4.93$& $1.19 $ & $19.7$ & 2.6 \\
I4-WRWR & WENO & WENO & $10^4$ & $3.125\times 10^{0}$  & $14.3$ & $4.90$ & $6.54$& $1.40 $ & $2.54$ & 3.2 \\
I5-WRWR & WENO & WENO & $10^5$ & $3.125\times 10^{1}$  & $27.6$ & $7.38$ & $12.0$& $1.97 $ & $0.447$ & 4.3 \\
I6-WRWR & WENO & WENO & $10^6$ & $3.125\times 10^{2}$  & $196$ & $31.0 $& $70.4$& $8.95 $ & $0.253$ & 4.9 \\
I2-WRZR & WENO & WZMP & $10^2$ & $3.125\times 10^{-2}$ & $6.60$ & $1.73$ & $2.34$& $0.553 $ & $93.4$ & 3.6 \\
I3-WRZR & WENO & WZMP & $10^3$ & $3.125\times 10^{-1}$ & $14.3$ & $4.60$ & $6.11$& $1.41 $ & $24.2$ & 2.9 \\
I4-WRZR & WENO & WZMP & $10^4$ & $3.125\times 10^{0}$  & $21.3$ & $6.31$ & $9.11$& $1.76 $ & $3.47$ & 4.1 \\
I5-WRZR & WENO & WZMP & $10^5$ & $3.125\times 10^{1}$  & $42.7$ & $10.5$ & $17.8$& $2.79 $ & $0.659$ & 6.7 \\
I6-WRZR & WENO & WZMP & $10^6$ & $3.125\times 10^{2}$  & $354$ & $53.9$ & $126$& $17.2 $ & $0.457$ & 6.8 \\
I3-ZRZR & WZMP & WZMP & $10^3$ & $3.125\times 10^{-1}$ & $16.0$ & $5.49$ & $6.82$& $1.59 $ & $27.3$ & 2.8 \\
I4-ZRZR & WZMP & WZMP & $10^4$ & $3.125\times 10^{0}$  & $22.3$ & $7.13$ & $9.66$& $1.81 $ & $3.69$ & 3.3 \\
I5-ZRZR & WZMP & WZMP & $10^5$ & $3.125\times 10^{1}$  & $21.7$ & $6.58$ & $9.71$& $1.63 $ & $0.364$ & 5.6 \\
I6-ZRZR & WZMP & WZMP & $10^6$ & $3.125\times 10^{2}$  & $53.6$ & $12.6$ & $21.9$& $3.26 $ & $0.0806$ & 6.9 \\

\bottomrule
\end{tabular}
}
\end{table}
\end{center}

\clearpage
\begin{center}
\begin{table}[htbp]
\caption{High resolution ideal MHD simulations of the MRI.}
\label{tab:sim_table_highideal}
\scalebox{0.5}{
\begin{tabular}{ccccccccccc}
\toprule
(1) Model & (2) R1 & (3) R2 & (4) $\beta$ & (5) $P_0$ & (6) $E_{M} \times 10^{2}$ & (7) $E_{K} \times 10^{2}$ & (8) $w_{M} \times 10^{2}$ & (9) $w_{R} \times 10^{2}$ & (10) $\alpha \times 10^{2}$ & (11) $Pr_{m,{\rm num}}$ \\ \midrule
H2-WRWH & HLLR & HLL  & $10^2$ & $3.125\times 10^{-2}$ & $5.27$ & $1.65$ & $1.87$& $0.463$ & $74.9$ & 0.53 \\
H3-WRWH & HLLR & HLL  & $10^3$ & $3.125\times 10^{-1}$ & $8.77$ & $3.46$ & $3.69$& $0.913$ & $14.8$ & 0.41 \\
H4-WRWH & HLLR & HLL  & $10^4$ & $3.125\times 10^{0}$  & $6.70$ & $3.13$ & $3.04$& $0.763$ & $1.22$ & 0.49 \\
H5-WRWH & HLLR & HLL  & $10^5$ & $3.125\times 10^{1}$  & $8.04$ & $3.23$ & $3.62$& $0.747$ & $0.14$ & 0.37 \\
H6-WRWH & HLLR & HLL  & $10^6$ & $3.125\times 10^{2}$  & $5.34$ & $2.12$ & $2.31$& $0.531$ & $0.00913$ & 0.31 \\
H2-WHWH & HLL  & HLL  & $10^2$ & $3.125\times 10^{-2}$ & $4.88$ & $1.47$ & $1.76$& $0.409$ & $69.6$ & 0.95 \\
H3-WHWH & HLL  & HLL  & $10^3$ & $3.125\times 10^{-1}$ & $9.57$ & $3.54$ & $4.11$& $0.929$ & $16.1$ & 1.1 \\
H4-WHWH & HLL  & HLL  & $10^4$ & $3.125\times 10^{0}$  & $9.83$ & $3.93$ & $4.57$& $0.969$ & $1.77$ & 0.99 \\
H5-WHWH & HLL  & HLL  & $10^5$ & $3.125\times 10^{1}$  & $9.82$ & $3.31$ & $4.57$& $0.764$ & $0.171$ & 0.95 \\
H6-WHWH & HLL  & HLL  & $10^6$ & $3.125\times 10^{2}$  & $7.41$ & $2.32$ & $3.32$& $0.573$ & $0.0125$ & 0.82 \\
H2-WRWR & HLLR & HLLR & $10^2$ & $3.125\times 10^{-2}$ & $7.27$ & $2.15$ & $2.59$& $0.628$ & $103$ & 1.9 \\
H3-WRWR & HLLR & HLLR & $10^3$ & $3.125\times 10^{-1}$ & $15.8$ & $5.30$ & $6.54$& $1.40$ & $25.5$ & 1.2 \\
H4-WRWR & HLLR & HLLR & $10^4$ & $3.125\times 10^{0}$  & $24.5$ & $7.69$ & $10.4$& $1.89$ & $3.94$ & 1.9 \\
H5-WRWR & HLLR & HLLR & $10^5$ & $3.125\times 10^{1}$  & $27.4$ & $7.58$ & $12.0$& $1.75$ & $0.439$ &2.3  \\
H6-WRWR & HLLR & HLLR & $10^6$ & $3.125\times 10^{2}$  & $133$ & $24.8$ & $48.8$& $7.31$ & $0.179$ & 2.9 \\
\bottomrule
\end{tabular}
}
\end{table}
\end{center}

\clearpage
\begin{center}
\begin{table}[htbp]
\caption{Visco-resistive MHD simulations of the MRI.}
\label{tab:sim_table_04}
\scalebox{0.5}{
\begin{tabular}{ccccccccccc}
\toprule
(1) Model & (2) $\nu \times 10^3$ & (3) $\eta \times 10^3$ & (4) $\beta$ & (5) $P_0$ & (6) $E_{M} \times 10^{2}$ & (7) $E_{K} \times 10^{2}$ & (8) $w_{M} \times 10^{2}$ & (9) $w_{R} \times 10^{2}$ & (10) $\alpha \times 10^{2}$ & (11) $Pr_{m}$ \\ \midrule

V3-0-1 & 0.5 & 0.5 & $10^3$ & $3.125\times 10^{-1}$ & $10.3$ & $3.76$ & $4.37$& $0.917$ & $16.9$ & 1.0 \\
V4-0-1 & 0.5 & 0.5 & $10^4$ & $3.125\times 10^{0}$ & $12.6$ & $4.18$ & $5.69$& $1.12$ & $2.18$ & 1.0 \\
V5-0-1 & 0.5 & 0.5 & $10^5$ & $3.125\times 10^{1}$ & $12.5$ & $3.82$ & $5.82$& $0.979$ & $0.218$ & 1.0 \\
V6-0-1 & 0.5 & 0.5 & $10^6$ & $3.125\times 10^{2}$ & $16.1$ & $4.53$ & $7.37$& $1.21$ & $0.0274$ & 1.0 \\
V3-1-1 & 1.0 & 1.0 & $10^3$ & $3.125\times 10^{-1}$ & $8.86$ & $4.79$ & $3.65$& $0.719$ & $14.0$ & 1.0 \\
V4-1-1 & 1.0 & 1.0 & $10^4$ & $3.125\times 10^{0}$ & $9.19$ & $3.04$ & $4.17$& $0.830$ & $1.60$ & 1.0 \\
V5-1-1 & 1.0 & 1.0 & $10^5$ & $3.125\times 10^{1}$ & $11.2$ & $3.13$ & $5.13$& $0.803$ & $0.190$ & 1.0 \\
V6-1-1 & 1.0 & 1.0 & $10^6$ & $3.125\times 10^{2}$ & $11.5$ & $3.26$ & $5.30$& $0.859$ & $0.0197$ & 1.0 \\
V3-2-1 & 2.0 & 2.0 & $10^3$ & $3.125\times 10^{-1}$ & $12.4$ & $3.70$ & $4.55$& $0.611$ & $16.5$ & 1.0 \\
V4-2-1 & 2.0 & 2.0 & $10^4$ & $3.125\times 10^{0}$ & $12.3$ & $2.91$ & $5.08$& $0.822$ & $1.89$ & 1.0 \\
V5-2-1 & 2.0 & 2.0 & $10^5$ & $3.125\times 10^{1}$ & $16.7$ & $3.46$ & $6.79$& $0.847$ & $0.244$ & 1.0 \\
V6-2-1 & 2.0 & 2.0 & $10^6$ & $3.125\times 10^{2}$ & $16.0$ & $3.87$ & $6.48$& $0.829$ & $0.0234$ & 1.0 \\
V3-4-1 & 4.0 & 4.0 & $10^3$ & $3.125\times 10^{-1}$ & $26.3$ & $3.03$ & $8.26$& $0.687$ & $28.6$ & 1.0 \\
V4-4-1 & 4.0 & 4.0 & $10^4$ & $3.125\times 10^{0}$ & $74.2$ & $7.41$ & $23.6$& $1.72$ & $8.10$ & 1.0 \\
V5-4-1 & 4.0 & 4.0 & $10^5$ & $3.125\times 10^{1}$ & $216.0$ & $22.9$ & $73.2$& $6.33$ & $2.55$ & 1.0 \\
V6-4-1 & 4.0 & 4.0 & $10^6$ & $3.125\times 10^{2}$ & $358.0$ & $33.7$ & $117.0$& $9.41$ & $0.403$ & 1.0 \\
\midrule
V3-2-2 & 2.0 & 1.0 & $10^3$ & $3.125\times 10^{-1}$ & $14.9$ & $3.55$ & $5.80$& $0.907$ & $21.5$ & 2.0 \\
V4-2-2 & 2.0 & 1.0 & $10^4$ & $3.125\times 10^{0}$ & $12.7$ & $4.04$ & $5.49$& $0.935$ & $2.06$ & 2.0 \\
V5-2-2 & 2.0 & 1.0 & $10^5$ & $3.125\times 10^{1}$ & $19.7$ & $4.25$ & $8.67$& $1.19$ & $0.316$ & 2.0 \\
V6-2-2 & 2.0 & 1.0 & $10^6$ & $3.125\times 10^{2}$ & $14.8$ & $3.12$ & $6.41$& $0.877$ & $0.0233$ & 2.0 \\
V3-4-2 & 4.0 & 2.0 & $10^3$ & $3.125\times 10^{-1}$ & $26.4$ & $3.35$ & $8.69$& $0.924$ & $30.8$ & 2.0 \\
V4-4-2 & 4.0 & 2.0 & $10^4$ & $3.125\times 10^{0}$ & $48.4$ & $8.41$ & $17.3$& $2.12$ & $6.22$ & 2.0 \\
V5-4-2 & 4.0 & 2.0 & $10^5$ & $3.125\times 10^{1}$ & $123.0$ & $15.4$ & $41.3$& $4.32$ & $1.46$ & 2.0 \\
V6-4-2 & 4.0 & 2.0 & $10^6$ & $3.125\times 10^{2}$ & $598.0$ & $50.8$ & $188.0$& $12.8$ & $0.642$ & 2.0 \\
V3-2-4 & 2.0 & 0.5 & $10^3$ & $3.125\times 10^{-1}$ & $21.0$ & $9.72$ & $8.38$& $1.39$ & $31.3$ & 4.0 \\
V4-2-4 & 2.0 & 0.5 & $10^4$ & $3.125\times 10^{0}$ & $43.9$ & $10.4$ & $18.5$& $3.05$ & $6.90$ & 4.0 \\
V5-2-4 & 2.0 & 0.5 & $10^5$ & $3.125\times 10^{1}$ & $35.7$ & $7.07$ & $15.3$& $1.95$ & $0.551$ & 4.0 \\
V6-2-4 & 2.0 & 0.5 & $10^6$ & $3.125\times 10^{2}$ & $76.6$ & $12.8$ & $30.1$& $3.90$ & $0.109$ & 4.0 \\
V3-4-4 & 4.0 & 1.0 & $10^3$ & $3.125\times 10^{-1}$ & $28.7$ & $6.31$ & $10.1$& $1.36$ & $36.8$ & 4.0 \\
V4-4-4 & 4.0 & 1.0 & $10^4$ & $3.125\times 10^{0}$ & $50.4$ & $8.20$ & $19.0$& $2.45$ & $6.86$ & 4.0 \\
V5-4-4 & 4.0 & 1.0 & $10^5$ & $3.125\times 10^{1}$ & $198.0$ & $23.9$ & $65.2$& $6.99$ & $2.31$ & 4.0 \\
V6-4-4 & 4.0 & 1.0 & $10^6$ & $3.125\times 10^{2}$ & $585.0$ & $48.0$ & $183.0$& $14.8$ & $0.633$ & 4.0 \\
\midrule
V3-1-0 & 1.0 & 2.0 & $10^3$ & $3.125\times 10^{-1}$ & $7.87$ & $3.34$ & $3.04$& $0.508$ & $11.4$ & 0.5 \\
V4-1-0 & 1.0 & 2.0 & $10^4$ & $3.125\times 10^{0}$ & $8.62$ & $2.33$ & $3.52$& $0.573$ & $1.31$ & 0.5 \\
V5-1-0 & 1.0 & 2.0 & $10^5$ & $3.125\times 10^{1}$ & $7.96$ & $2.50$ & $3.44$& $0.517$ & $0.127$ & 0.5 \\
V6-1-0 & 1.0 & 2.0 & $10^6$ & $3.125\times 10^{2}$ & $8.51$ & $2.09$ & $3.49$& $0.501$ & $0.0128$ & 0.5 \\

\bottomrule
\end{tabular}
}
\end{table}
\end{center}

\end{document}